\documentclass[%
 aip,
 amsmath,amssymb,
preprint,%
]{revtex4-1}

\usepackage{graphicx}
\usepackage{dcolumn}
\usepackage{bm}

\usepackage[utf8]{inputenc}
\usepackage[T1]{fontenc}
\usepackage{mathptmx}
\usepackage{etoolbox}
\usepackage{mhchem}
\usepackage{graphics}
\usepackage{booktabs}%

\usepackage[utf8]{inputenc}
\usepackage[T1]{fontenc}
\usepackage{mathptmx}
\usepackage{etoolbox}
\usepackage{mhchem}
\usepackage{graphics}
\usepackage{booktabs}%
\usepackage{xcolor}%

\usepackage{listings}
\lstdefinestyle{shared}
{
    numbers=left,
    numbersep=1em,
    numberstyle=\tiny\color{black},
    frame=single,
    framesep=\fboxsep,
    framerule=\fboxrule,
    rulecolor=\color{black},
    xleftmargin=\dimexpr\fboxsep+\fboxrule\relax,
    xrightmargin=\dimexpr\fboxsep+\fboxrule\relax,
    breaklines=true,
    tabsize=2,
    columns=flexible,
}
\lstdefinestyle{plumed}
{
    style=shared,
    language={bash},
    basicstyle=\scriptsize\tt,
    keywordstyle=\color{darkgray},
    commentstyle=\color{gray},
    backgroundcolor=\color{gray!10},
    morekeywords={
    COORDINATIONNUMBER, MATHEVAL, SPECIESA, SPECIESB, SWITCH, MEAN, ARG, LOCAL_Q6, Q6, FUNC, PERIODIC,
    ENVIRONMENTSIMILARITY, SPECIES, SIGMA, CRYSTAL_STRUCTURE, MORE_THAN, LOWMEM, VMEAN,
    REFERENCE, REFERENCE_1, REFERENCE_2, REFERENCE_3, REFERENCE_4, REFERENCE_5, REFERENCE_6
    },
}
\lstnewenvironment{plumed}
{\lstset{style=plumed}
 \lstset{escapeinside={(*@}{@*)}}}
{}

\NewDocumentCommand{\MakeTitleInner}{ +m +m +m }{
    \newpage%
    \null%
    \vskip 2em%
    \begin{center}%
        \let \footnote \thanks
        {\LARGE #1 \par}
        \vskip 1.5em%
        {%
            \large
            \lineskip .5em%
            \begin{tabular}[t]{c}%
                #2
            \end{tabular}\par%
        }%
        \vskip 1em%
        {\large #3}
    \end{center}%
    \par
    \vskip 1.5em%
}
\NewDocumentCommand{\MakeTitle}{ +m +m +m }{%
    \begingroup
        \renewcommand\thefootnote{\@fnsymbol\c@footnote}%
        \def\@makefnmark{\rlap{\@textsuperscript{\normalfont\@thefnmark}}}%
        \long\def\@makefntext##1{\parindent 1em\noindent
            \hb@xt@1.8em{%
                \hss\@textsuperscript{\normalfont\@thefnmark}%
            }##1%
        }%
        \if@twocolumn
            \ifnum \col@number=\@ne
                \MakeTitleInner{#1}{#2}{#3}
            \else
                \twocolumn[\MakeTitleInner{#1}{#2}{#3}]%
            \fi
        \else
            \newpage
            \MakeTitleInner{#1}{#2}{#3}
        \fi
        \thispagestyle{plain}
    \endgroup
    \setcounter{footnote}{0}%
}
\makeatother

\makeatletter
\def\@email#1#2{%
 \endgroup
 \patchcmd{\titleblock@produce}
  {\frontmatter@RRAPformat}
  {\frontmatter@RRAPformat{\produce@RRAP{*#1\href{mailto:#2}{#2}}}\frontmatter@RRAPformat}
  {}{}
}%
\makeatother
\begin{document}

\preprint{}

\title[]{Phase Diagram of ZIF-4 Computed via Well-tempered Metadynamics}

\author{Emilio Mendez}
\author{Rocio Semino}%
 \email{rocio.semino@sorbonne-universite.fr}
\affiliation{ 
Sorbonne Université, CNRS, Physico-chimie des Electrolytes et Nanosystèmes Interfaciaux, PHENIX, F-75005 Paris, France
}%

\date{\today}%

\begin{abstract}
Well-tempered metadynamics simulations are employed to explore the phase diagram of ZIF-4, a porous crystalline metal-organic framework of industrial relevance. Despite the vast amount of experimental efforts, the phase diagram that includes ZIF-4 and its related polymorphs has not yet been fully determined. For example, the crystalline phase called ZIF-4-cp is not experimentally observed when high pressure ramps are applied. Our simulations shed light into the phase diagram topology and allow us to further look into the collective degrees of freedom that drive the phase transitions in the T=150-450 K and P=0-200 MPa region. The porous ZIF-4 phase transforms into ZIF-4-cp through pore closure, while the latter has a phase transition at higher pressures regimes to ZIF-4-cp-II, a transformation which involves subtle changes in swing dihedral angles.

\end{abstract}

\maketitle

\section{\label{sec1}Introduction}

Exploring the phase diagram of metal-organic frameworks (MOFs), porous materials that can act as hosts with  sieving, confining and nanoreactor properties, is not an easy task. This is particularly true for ZIF-4,\cite{Park2006} a porous crystalline orthorhombic MOF composed of Zn$^{2+}$ cations and imidazolate molecules as ligands that adopts the \textit{cag} topology. This MOF and its complex phase diagram, featuring at least six crystalline and two amorphous polymorphs,\cite{Widmer2019} has been puzzling scientists for more than one decade at the moment of this writing.\cite{Bennett2010} 

Countless reasons explain the hype surrounding ZIF-4. First, its capacity to transform into porous liquids\cite{Gaillac2017} and amorphous polymorphs, which has allowed scientists to establish the basis of MOF-based glass technologies\cite{Bennett2015,Bennett2018,Madsen2020,Wei2023,Kim2024, Xue2024} promising to ally hydrocarbon separation properties\cite{Hartmann2015, Hovestadt2018} with easy shaping\cite{BazerBachi2014} and scaled-up synthesis\cite{Hovestadt2017} for industrial applications.  Moreover, changing the composition of ZIF-4 by substituting imidazole by other ligands helps tune its melting temperature and adsorption properties.\cite{FrentzelBeyme2019,Yu2020} From the applications point of view, ZIF-4 is a stimuli-responsive material,\cite{Iacomi2021} that changes its structure when subjected to changes in temperature,\cite{Bennett2011} pressure,\cite{Bennett2011_3,Henke2018,Vervoorts2019} x-ray absorption\cite{Widmer2019_2} and guest adsorption,\cite{Henke2018} among others. Stimuli responsive properties have made of ZIF-4 an excellent candidate for reversible trapping of potentially harmful guests.\cite{Beake2013} Finally, ZIF-4 also exhibits auxetic behaviour, as demonstrated both by a variety of experimental\cite{Ryder2014,Butler2019}, \textit{ab initio}\cite{Tan2015,Ryder2016}, classical and reactive modelling studies.\cite{BousselduBourg2014,Castel2023_2} 

A better understanding of the complex phase diagram of ZIF-4 is required to master its properties and help further develop the MOF-glasses field.
The challenges of this task range from glass transition temperature values that are sensible to changes in synthesis conditions,\cite{Zhang2019} to the still ongoing experimental developments that are needed to overcome slow kinetics phase transformations.\cite{Gong2022}
Indeed, the detection of high pressure/temperature phases is strongly dependent on pressure/temperature ramps. This is how Widmer and collaborators\cite{Widmer2019} explain why they do not observe the ZIF-4-cp polymorph, whose existence was previously demonstrated by Henke and coworkers,\cite{Henke2018} in their phase diagram determination, that is the most complete up to date. Indeed, upon an increase of pressure, the authors observed a phase transition from ZIF-4 to ZIF-4-cp-II, a more compact polymorph than ZIF-4-cp ($V_{ZIF-4} = 4342 \AA^3/UC$, $V_{ZIF-4-cp-II} = 3055 \AA^3/UC$ and $V_{ZIF-4-cp} = 3457 \AA^3/UC$). The authors speculate that ZIF-4-cp must lie in between ZIF-4 and ZIF-4-cp-II in the phase diagram,\cite{Widmer2019,Widmer2019Nature} but they could not observe it. At the moment of this writing, the place of ZIF-4-cp in the phase diagram is thus unknown, let alone the sign of the Clapeyron lines that determine its stability region.

Furthermore, detecting and measuring the collective motions that take place at the molecular level and drive these phase transitions forward is practically impossible at the experimental level. Molecular simulation allows us to access molecular detail that cannot yet be measured, filling in this gap. In this spirit, we have recently modeled the ZIF-4 $\longrightarrow$ ZIF$\_$a and ZIF-zni $\longrightarrow$ ZIF$\_$liq phase transitions via molecular dynamics simulations\cite{Mendez2024} relying on a force field\cite{Balestra2022} that partially incorporates reactivity through the treatment of coordination bonds as non-bonded interactions. Three machine learning based force fields were recently developed for ZIF-4 and applied to study its thermal phase transitions. Shi and coworkers\cite{Shi2024} and Du et al\cite{Du2024} have relied on deep-learning potentials while Castel and coworkers chose message-passing neural networks to model glassy MOFs.\cite{Castel2024} These force fields strive for keeping \textit{ab initio} accuracy while tackling larger systems (for instance, cell sizes compatible with the amorphous phases) than those that were previously accessible by first principles calculations.\cite{Gaillac2018} Our nb-ZIF-FF force field\cite{Balestra2022} is the only reactive physically-motivated force field to date that was successful in accurately modeling amorphous ZIFs and their crystalline polymorphs, since efforts to use ReaxFF for this task have been controversial.\cite{Yang2018,Castel2022} In the works by Méndez and Shi,\cite{Mendez2024,Shi2024} a lower-density amorphous phase, which had been experimentally observed as an intermediate between ZIF-4 and ZIF$\_$a,\cite{Bennett2015} was also detected. Méndez and Du\cite{Mendez2024,Du2024} both detected changes in the coordination bonds underlying the amorphization process. On a related note, Sapnik and collaborators have addressed the important problem of generating representative configurations for ZIF$\_$a via an \textit{in silico} polymerization procedure.\cite{Sapnik2021}

Despite the intense activity in the field, no modeling study has yet tackled the exploration of the ZIF-4 phase diagram with varying pressure up to date. This is due to the fact that in order to explore pressure changes it is not enough to have a high quality reactive force field, but it is also required to couple it with advanced enhanced sampling methods\cite{Stracke2024} that allow overcoming free energy barriers much larger than $k_b T$ around ambient temperature conditions. 

In this work, we combine a reactive force field together with well-tempered metadynamics simulations to explore pressure induced phase transitions undergone by ZIF-4 around ambient temperature. We provide for the first time a computed topology for the phase diagram including ZIF-4-cp, the phase that was missing in the experimental studies.\cite{Widmer2019} We reveal the signs of the Clapeyron lines between ZIF-4, ZIF-4-cp, ZIF-4-cp-II and ZIF$\_$a. Furthermore, we identify the collective modes at the onset of the phase transitions at the molecular level, which are related to pore closure for the ZIF-4 $\longrightarrow$ ZIF-4-cp transition and to swing ligand rotations for the ZIF-4-cp $\longrightarrow$ ZIF-4-cp-II one.     

\section{Methods}\label{methods}

\subsection{Metadynamics}\label{metadynamics}

A trajectory in which all the species of interest are visited multiple times is needed to accurately study the relative stability of the different phases at a fixed thermodynamic condition. Nevertheless, the inter-conversions between polymorphs involve the crossing of high activation energy barriers, making it necessary to implement an enhanced sampling scheme to reduce the time required to collect enough statistics.
The metadynamics\cite{Laio2002} procedure is a biased 
enhanced sampling method that consists in dynamically adding energy
penalties in the potential energy surface regions that are visited more frequently. 
This allows the system to efficiently explore the whole collective variable space and surpass free energy barriers that would have taken times several orders of magnitude higher to overcome in standard unbiased molecular dynamics simulations. 

An advantage of metadynamics with respect to other enhanced sampling procedures is that the free energy surface as a function of the collective variables can be recovered directly from the negative of the bias potential together with a reweighting procedure,\cite{Schfer2020} without the need of extra calculations.
The well-tempered\cite{Barducci2008} version of the technique allows a better convergence by diminishing the rate at which the bias potential evolves as the simulation progresses.

\subsection{Multibaric ensemble}\label{multibaric}

A key aspect of metadynamics simulations is the adequate selection of the order parameters to bias,
as this will determine the extent to which the mechanistic and energetic information obtained can be correlated to the physical phenomena under investigation.\cite{Bussi2020}
A common practice in processes that involve phase transitions is to use the systems' energy and/or volume as collective variables, since these quantities tend to vary significantly between phases.
This choice presents an additional advantage: modeling the relative stability of two phases in the $(T,P)$ plane usually requires to perform the simulations at multiple thermodynamic conditions, which demands a large 
amount of computing time.
However, this choice of collective variables makes it possible to extrapolate results obtained at one specific thermodynamic state to another by the use of the so called multithermal multibaric ensemble formalism.\cite{Piaggi2019}
This method consists in the calculation of the free energy as a function of the potential energy $E$ and volume $V$ of the system, $G^{(T,P)}(E,V)$, at the temperature $T$ and pressure $P$ at which the simulation is performed.
Then, the free energy at a different state $(T',P')$ is recovered by the following relation: 
\begin{equation} \label{multitmultip}
    \beta' G^{(T',P')}(E,V) = \beta G^{(T,P)}(E,V) + (\beta' - \beta)E + (\beta'P' - \beta P)V + C
\end{equation}
where $\beta=1/k_bT$, $\beta'=1/k_bT'$, and $C$ is a constant that does not depend on $E$ and $V$. 

Since we aimed to study the properties of the polymorphs at different pressure conditions but at constant temperature, we did not bias the energy. For an isotherm, equation (\ref{multitmultip}) simplifies to:
\begin{equation} \label{multip}
    G^{(P')}(V) = G^{(P)}(V) + (P' - P)V + C 
\end{equation}

When we analyzed the phase transitions experienced by ZIF-4 under this setup, we noticed that the use of the volume alone as order parameter for metadynamics was not enough to efficiently sample all the possible states. 
The reason for this is that some of the polymorphs differ only slightly in their total volume, albeit they present large variations in their cell parameter ratios.
Therefore, we decided to use the three cell dimensions as collective variables
instead of the volume as done by Martoňák and collaborators.\cite{Martok2003} For this setup, equation (\ref{multip}) trivially turns into:
\begin{equation} \label{multicell}
    G^{(P')}(a,b,c) = G^{(P)}(a,b,c) + (P' - P)V + C 
\end{equation}
Where $a$, $b$ and $c$ are the dimensions of the simulation box in the $x$, $y$ and $z$ directions respectively, and $V=a.b.c$ since we only considered orthorhombic polymorphs. 
In order to check the robustness of the method, three simulations were carried out at different pressure conditions of 0, 40 and 80 MPa respectively. These values were chosen to lie in the pressure region where the phase transitions were found experimentally.\cite{Widmer2019,Henke2018}

\subsection{Simulation details}\label{details}

We performed the simulations through the LAMMPS open source software\cite{lammps}, coupled with PLUMED\cite{Tribello2014} package for implementing well-tempered metadynamics. The nb-ZIF-FF force field was used to model the interactions\cite{Balestra2022}. This force field has two main features: (i) the inclusion of dummy atoms in Zn$^{2+}$ and N (within imidazole) species to correctly reproduce the tetrahedral coordination environment around the metal, and (ii) the possibility of metal-ligand bond breaking/formation through the use of non bonded interaction terms in the form of Morse potentials. This force field is also known to reproduce the experimental properties of several ZIF polymorphs including ZIF-4, along with the corresponding glasses obtained through ambient pressure thermal amorphization.\cite{Balestra2022,Mendez2024} Bonded terms are based on the ZIF-FF force field\cite{Weng2019} and include harmonic style bonds, harmonic plus an additional Urey Bradley term for angles and cosine based dihedrals and impropers. Coulombic interactions are computed via the particle-particle/particle-mesh method while all dispersion interactions (Morse potentials for coordination bonds and 12-6 Lennard-Jones for the rest of the species) were computed considering a cutoff of 1.3 nm. 

The integration of the equations of motion was done in the NPT ensemble, using Nose-Hoover thermostats and barostats \cite{Evans1985}. The temperature was set to 300 K, and the damping parameters were set to 100 and 1000 time steps for regulating temperature and pressure respectively.
The barostat acted independently in each of the three system dimensions to allow the cell parameters to evolve in an unconstrained way. In all cases the simulation box was kept orthorhombic. 
The parameters for well-tempered metadynamics were chosen to be 9 kJ/mol and 0.1 nm for the initial gaussian height and standard deviation, respectively, and 75 for the bias factor that controls the decay of the heights with time.
Five parallel walkers were employed to accelerate the metadynamics convergence.\cite{Raiteri2005} Each walker evolves independently from the others but they all share the same bias potential obtained from the addition of gaussian terms.
The time step was set to 0.5 fs and the total time for each simulation, comprising all the walkers, was around 100 ns. 
Additional constraints were included for preventing the system from exploring non physical regions. These consisted in upper and lower bounds for each of the cell parameters and for the total volume, as well as a lower bound for the Zn-N total coordination to avoid metal-ligand bond breaking events that could lead to amorphization. This is because no amorphous phase has been experimentally reported in the phase diagram region that we target for exploration in this work. As a consequence, all the studied phases retained the connectivity of ZIF-4,
resulting in a tetra-coordination for all the Zn$^{2+}$ in the simulation box. 
Further details related to the well-tempered metadynamics simulations can be found in the Supplementary Information (SI).

Metadynamics simulations were conducted at 0, 40 and 80 MPa. All of them started from a 2x2x2 super cell of ZIF-4, which contains a total of 128 Zn$^{2+}$ atoms and has dimensions of 3.08x3.06x3.68 nm$^3$. 

Additionally, a few unbiased molecular dynamics simulations were performed for each of the obtained polymorphs. These were all carried out in the NPT ensemble for the thermodynamic conditions specified in each case, with the same force field, thermostat, barostat and timestep as those reported for the well-tempered metadynamics.

\section{Results}\label{sec2}

\subsection{Polymorph Stability}\label{stability}

We computed the free energy as a function of the cell parameters $G^{(P)}(a,b,c)$ from the metadynamics simulations at pressure $P=0$, 40 and 80 MPa at which the trajectories were obtained. The convergence criterion, as well as the technique employed for averaging the results and computing the errors, were based on the procedure developed by Tiwary \textit{et al.}\cite{Tiwary2014}, more details are provided in the SI.
From these initial free energy surfaces, we calculated $G^{(P')}(a,b,c)$ at different pressures $P'$ in the range that goes from 0 to 300 MPa via the equation (\ref{multicell}).
The difference between the three free energy surfaces at a given pressure was found to be on the same order of magnitude as the individual error bars ($\sim 0.5$ kJ/mol), which proves that the multibaric method is consistent. In what follows, results from the three simulations were averaged.

\begin{figure}[h]
\centering
\includegraphics[width=0.5\textwidth]{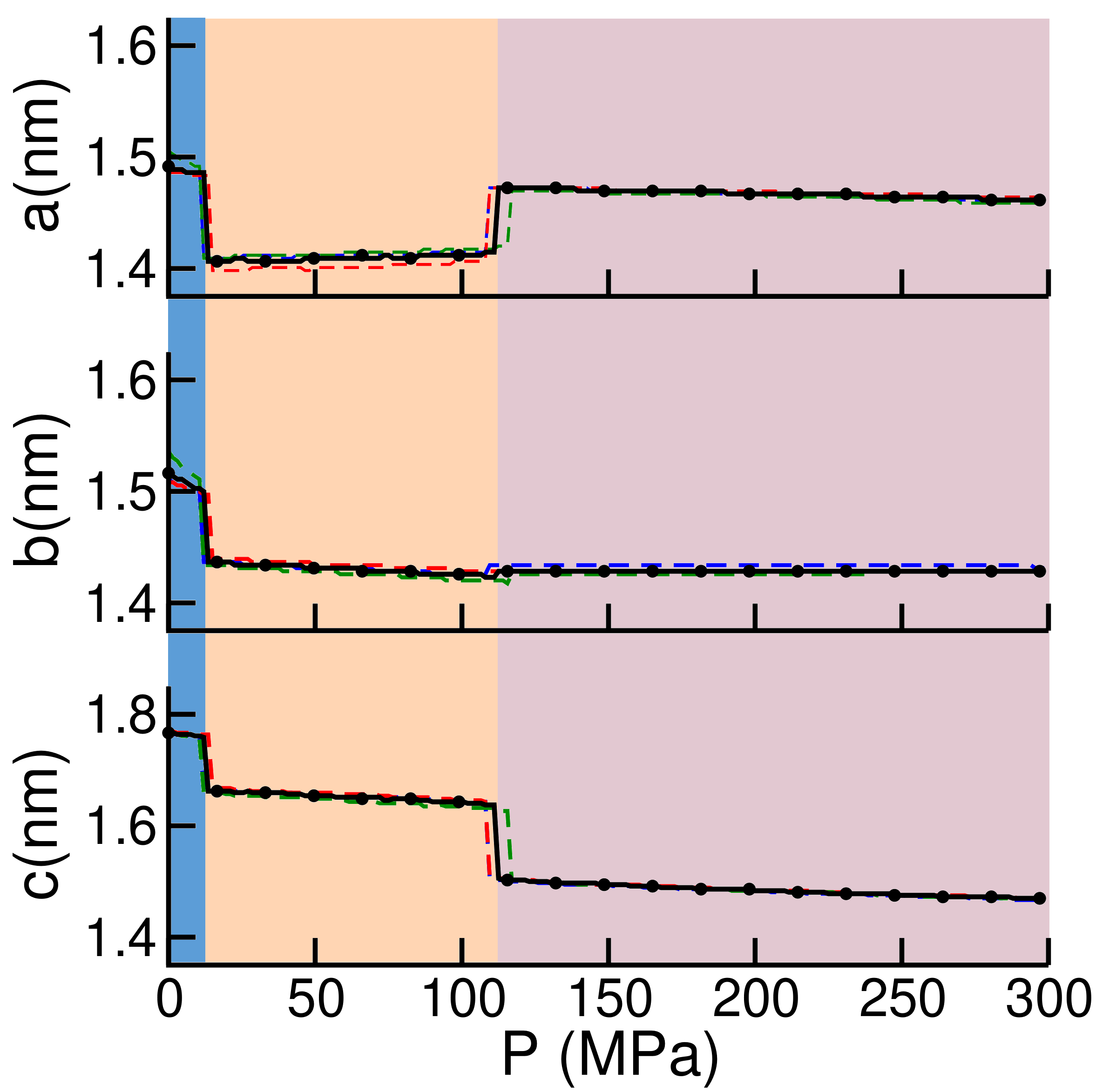}
\caption{\label{fig:cell}{Evolution of equilibrium cell parameters as a function of the pressure. The blue, red and green dashed lines correspond to results obtained from well-tempered metadynamics simulations at 0, 40 and 80 MPa respectively, while the black dotted line corresponds to the average. Regions are colored according to the most stable polymorph stability regions: blue corresponds to ZIF-4, orange to ZIF-4-cp and pink to ZIF-4-cp-II.}}
\end{figure}

We started our analysis by computing the position in the $(a,b,c)$ space of the absolute minimum of the free energy surface as a function of pressure. This minimum indicates the cell parameters of the most stable polymorph at the selected thermodynamic conditions. The results are summarized in Fig. \ref{fig:cell}.
We observe three pressure regions, separated by discontinuities in the lattice constants. These abrupt jumps correspond to the inversion of the relative stability of two distinct free energy minima, a characteristic of first order phase transitions. The curves obtained from individual simulations have the same qualitative behaviour, with only a slight difference in the absolute value of the  pressure of the transitions.
The first discontinuity, at $P=13 \pm 1$ MPa corresponds to a contraction of all the three cell parameters, which represents an overall change in the unit cell volume of -0.58 nm$^3$ (-15\%). The second transition takes place at $P=110 \pm 4$ MPa, and it features a contraction in $z$ direction while the $a$ and $b$ parameters slightly expand. As a result, the change in unit cell volume associated with this process was found to be of -0.13 nm$^3$ (-4\%).

By comparison of the cell parameters in each pressure region with the experimental lattice constants of several polymorphs\cite{Henke2018,Widmer2019}, we associated the first region to ZIF-4 (blue), the second one to ZIF-4-cp (orange), and the third one to ZIF-4-cp-II (pink). 
The lattice constants and volumes of the three species are summarized in table \ref{tabla}. The results are within a margin of error of less than 10\% in all cases with respect to the experimental values.

\begin{figure}[h]
\centering
\includegraphics[width=0.75\textwidth]{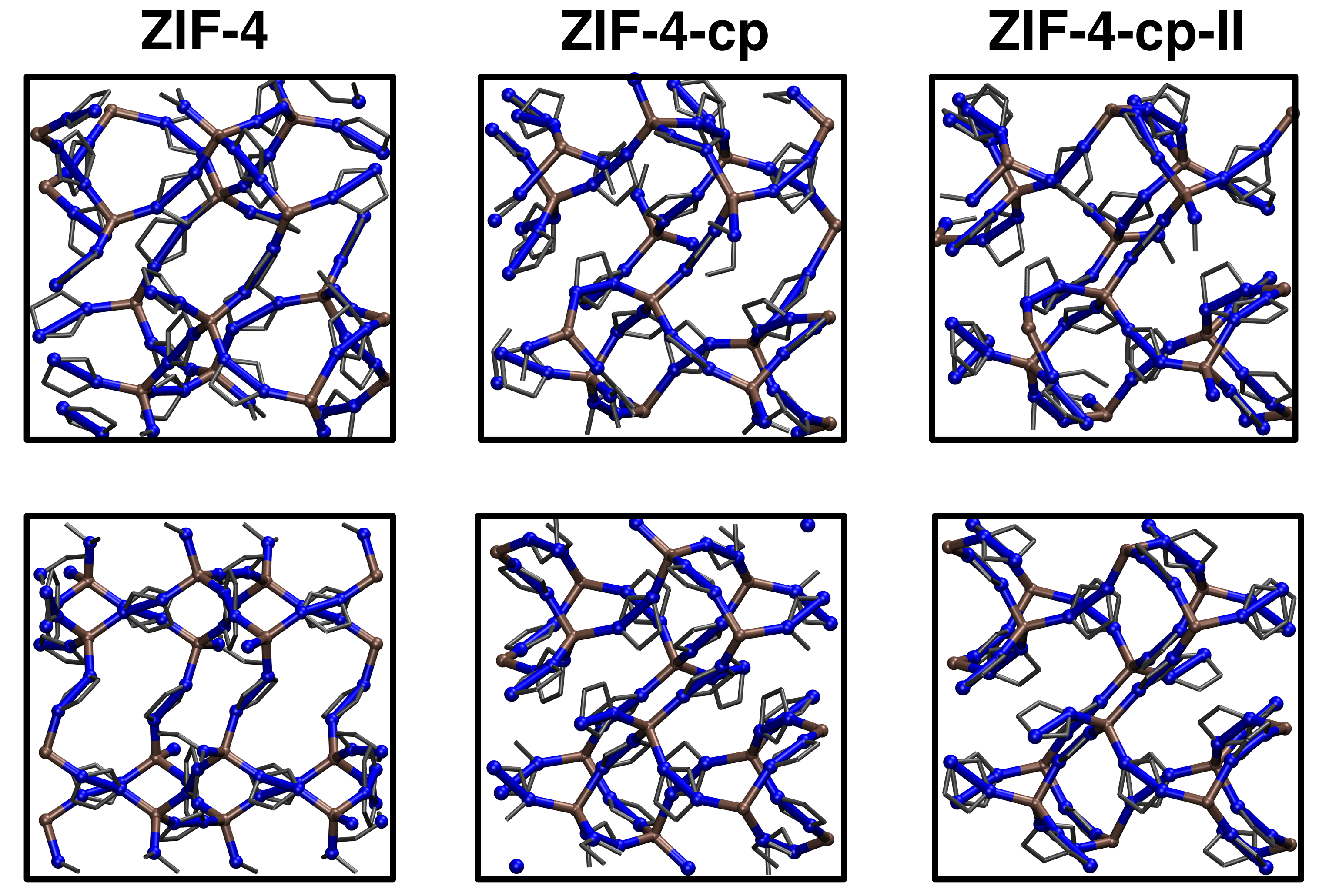}
\caption{\label{fig:polymorphs}{Representative snapshots of the studied ZIF-4 polymorphs in the $xy$ plane. The upper (lower) pictures correspond to the well-tempered metadynamics (experimental) structures. Color code: Zn (dark brown), N (blue), C (grey). For clarity purposes, H atoms were ignored and the snapshots have been scaled up to the same size.}}
\end{figure}

In Fig. \ref{fig:polymorphs} we present typical configurations of the polymorphs obtained from our simulations along with the reported experimental structures. The snapshots were obtained from individual simulations at pressure conditions of stability of each phase. The visual inspection of the high pressure polymorphs supports the previous assignment, taking into account that our structures are distorted by thermal fluctuations. Further analysis of the spatial arrangement of the obtained structures can be found in section \ref{microscopic}.

\begin{table}[h]
\caption{Cell parameters of the obtained polymorphs. Values in parenthesis correspond to experimental data\cite{Widmer2019}.}\label{tabla}%
\begin{tabular}{@{}llll@{}}
\toprule
               & ZIF-4  & ZIF-4-cp & ZIF-4-cp-II\\
\midrule
$P$(MPa)       & 0   & 80  & 150  \\
$V$(nm$^3$)    & 4.01 (4.34)  & 3.33 (3.46)  & 3.12 (3.25) \\
$a$(nm)        & 1.49 (1.54)  & 1.41 (1.42)  & 1.47 (1.46)   \\
$b$(nm)        & 1.52 (1.53)  & 1.43 (1.49)  & 1.43 (1.44) \\
$c$(nm)        & 1.77 (1.84)  & 1.65 (1.63)  & 1.49 (1.54)    \\
\botrule
\end{tabular}
\end{table}

As a consequence, the coexistence pressure between ZIF-4 and ZIF-4-cp was found to be $P\sim13$ MPa while that for the ZIF-4-cp/ZIF-4-cp-II pair transition it was $P\sim110$ MPa.
Experimentally, the ZIF-4/ZIF-4-cp equilibrium was registered to take place at $P\sim28$ MPa. This difference is inside the typical margin of error of calculations obtained using classical force fields.
The transition between ZIF-4-cp and ZIF-4-cp-II was never observed experimentally. Instead, the ZIF-4-cp-II polymorph was obtained directly from ZIF-4 at a pressure around $P\sim100$ MPa. As mentioned above, the authors argue that the lack of observation of the intermediate ZIF-4-cp phase is due to the high slope of the applied pressure ramp.\cite{Widmer2019} 
Our results support the affirmation that both high pressure polymorphs are stable under certain pressure ranges and they correspond to two distinct phases separated by a free energy barrier. 
The higher pressure polymorph ZIF-4-cp-III was not obtained in our simulations since we limited our analysis to orthorhombic unit cells, which is not the case for this polymorph. We relaxed this constraint to perform a simulation of isolated ZIF-4-cp to check if the $\beta$ cell parameter suffers a spontaneous distortion from 90$^{\circ}$ as stated in previous works\cite{Henke2018}, but no such deviation was found. This is in agreement with the experimental results from Vervoorts \textit{et al.}\cite{Vervoorts2019}
The high pressure amorphization of the system was also observed in preliminary simulations, but since we aimed to focus our analysis on crystalline structures, we constrained the connectivity of the system to avoid this phenomenon, as specified in the methods section.

The free energy surface obtained via the well-tempered metadynamics simulations is a 4D plot, since it depends on the value of the three collective variables $(a,b,c)$. In order to more easily visualize the obtained profile, we performed a dimensionality reduction from the original three variable output. To do so, we considered the free energy as a function of the unit cell volume, $G(V)$. The procedure to make this transformation is detailed in the SI.

\begin{figure}[h]
\centering
\includegraphics[width=0.5\textwidth]{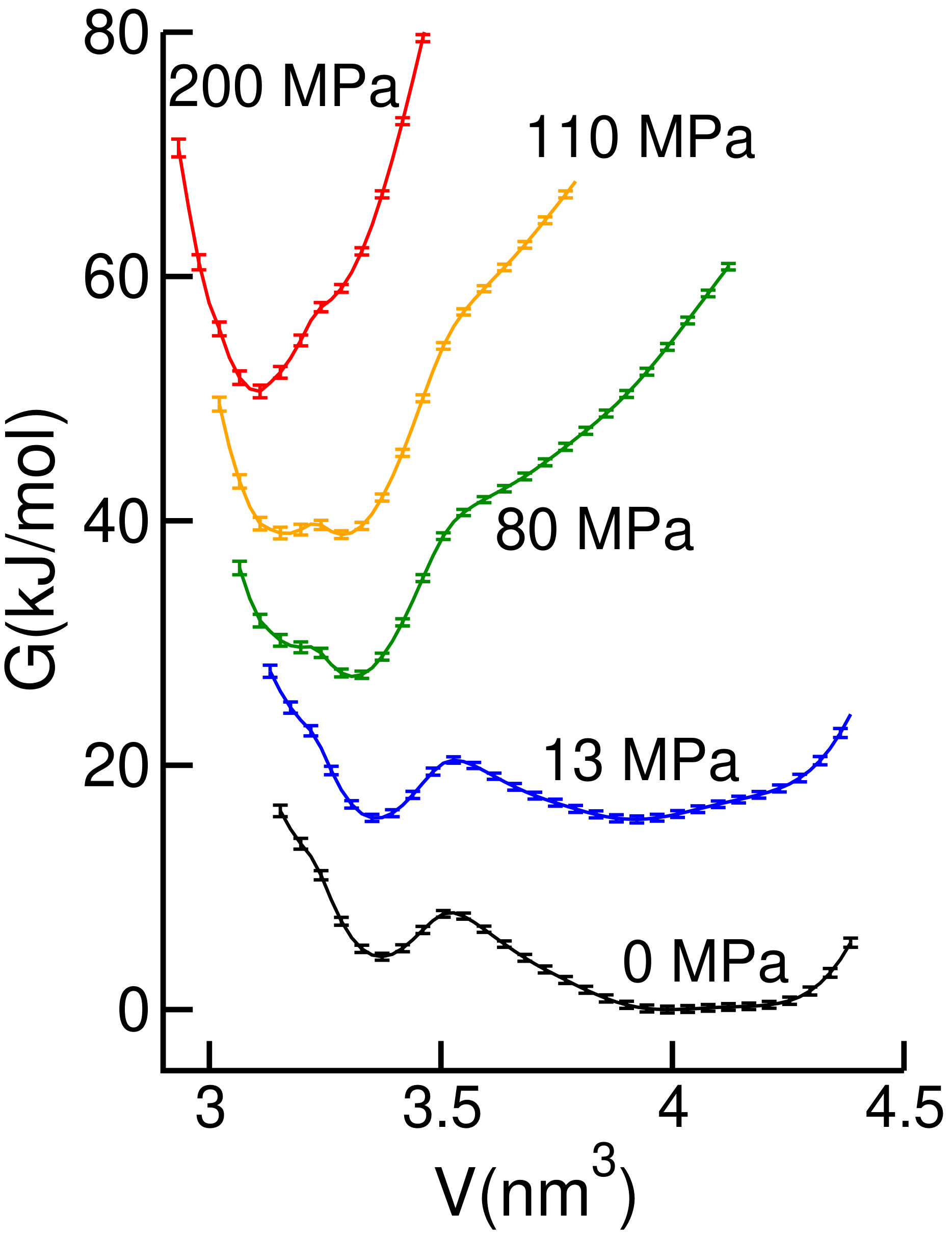}
\caption{\label{fig:gvp}{Free energy vs. volume  per unit cell at different pressures. The absolute height of each curve is arbitrary and it is adjusted to improve the clarity of the figure.}}
\end{figure}

\begin{figure}[h]
\centering
\includegraphics[width=0.5\textwidth]{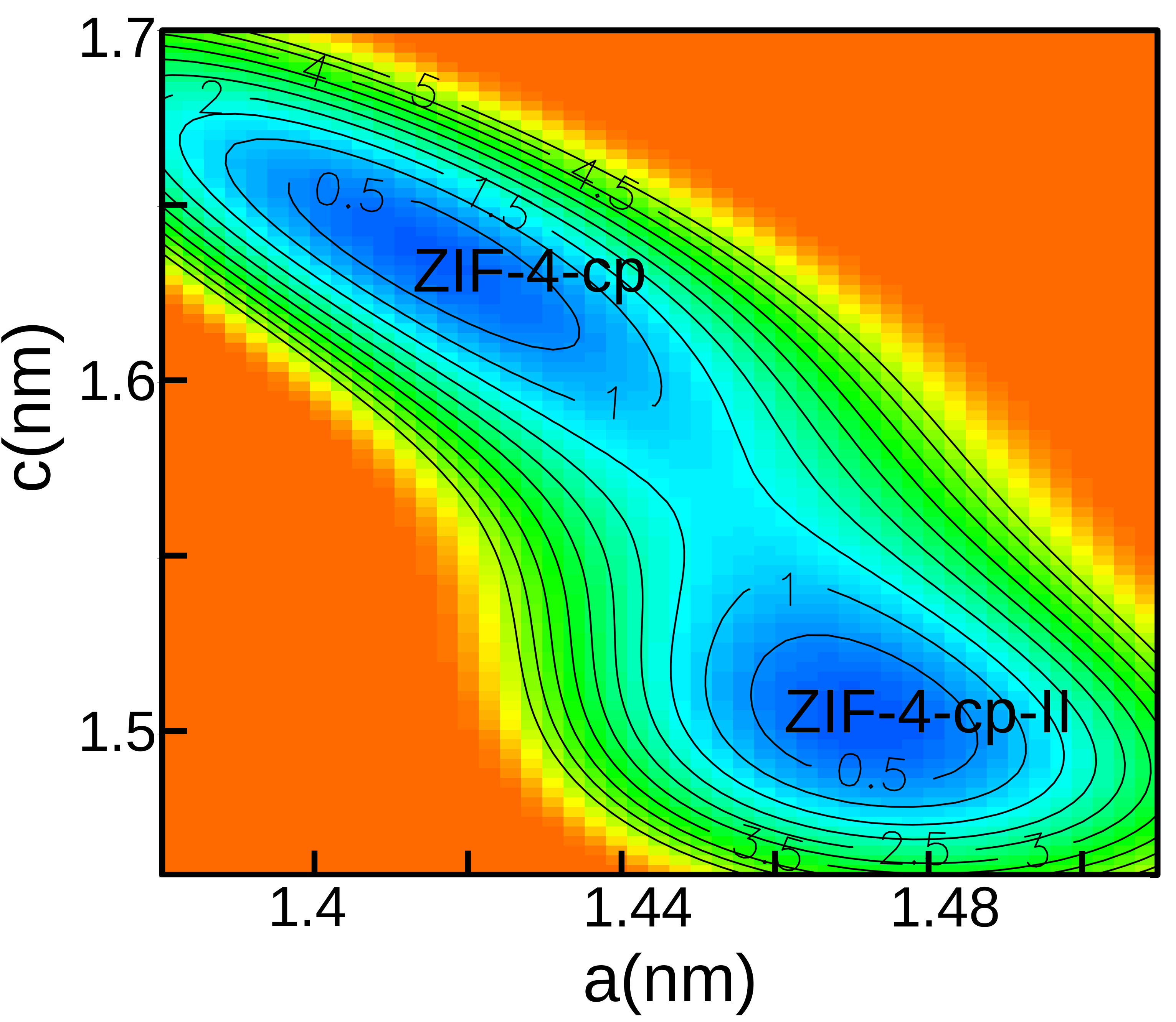}
\caption{\label{fig:2d}{Free energy as a function of cell parameters $a$ and $c$ at $P=110$ MPa (yellow curve in Fig. \ \ref{fig:gvp}). The zero of free energy is assigned to the minima, and the color bar goes from blue (lower energy) to red (higher energy). Isoenergy lines in kJ/mol are also shown. The values correspond to one unit cell.}}
\end{figure}

The result of $G(V)$ at different pressures is shown in Fig. \ref{fig:gvp}. 
The curve that corresponds to $P=0$ MPa -virtually ambient pressure- presents two minima. The lowest one at $V\sim4.0$ nm$^3$ can be assigned to ZIF-4 while the one at $V\sim3.4$ nm$^3$ corresponds to ZIF-4-cp. As expected, ZIF-4 is the most stable phase at these thermodynamic conditions.
At $P=13$ MPa, the two minima have the same height, meaning that the equilibrium pressure was achieved. The energy barrier that separates both states is in the order of 5 kJ/mol. After that, the ZIF-4-cp minimum becomes the most stable, as shown in the curve corresponding to $P=80$ MPa. 
At $P=110$ MPa we get to the ZIF-4-cp/ZIF-4-cp-II coexistence point. 
From this picture it is not clear if the free energy landscape can be described in terms of two minima, or just a single basin that presents large fluctuations.
To correctly characterize this transition, it is necessary to take into account not only the change in volume, but also the variation in the $a/c$ ratio, as previously explained. We thus plotted the free energy as a function of the cell parameters $a$ and $c$ in Fig.\ref{fig:2d}. This was done by integrating the $b$ coordinate from the original three dimensional profile. In this picture it is possible to confirm that there are two distinct phases, which correspond to different minima, separated by a free energy barrier on the order of 2 kJ/mol. As we discuss in section \ref{microscopic}, the lower barrier is due to the fact that the structural changes that take place in this phase transition are much more subtle than those associated to the ZIF-4/ZIF-4-cp phase transition. 
Finally, at higher pressures ZIF-4-cp-II becomes the most stable polymorph, as shown in the curve corresponding to $P=200$ MPa in Fig. \ref{fig:gvp}.

\subsection{Microscopic characterization of the phase transitions}\label{microscopic}

\begin{figure}[h]
\centering
\includegraphics[width=0.4\textwidth]{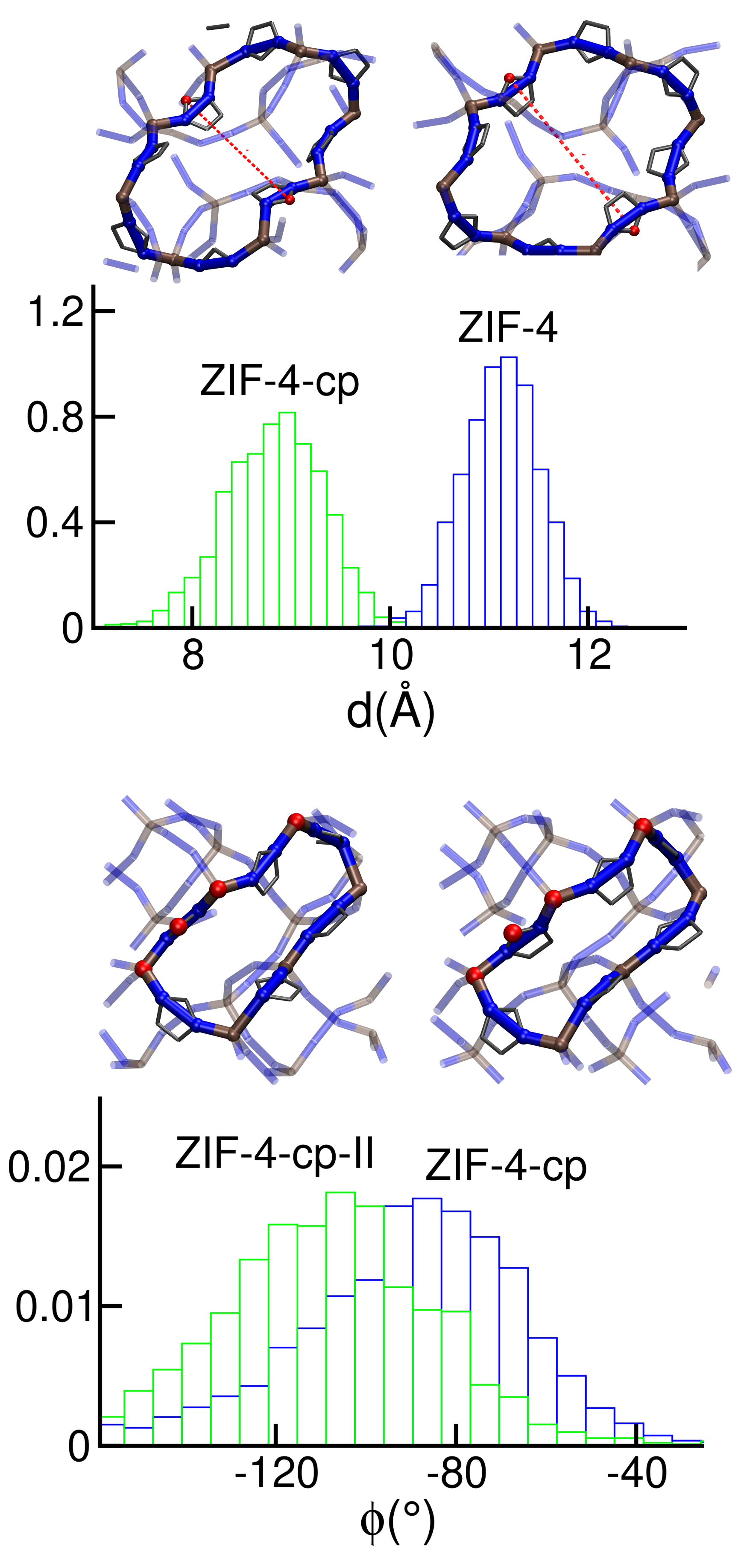}
\caption{\label{fig:histograms}{Top: comparison between snapshots of ZIF-4-cp (left) and ZIF-4 (right) 8-membered ring pore windows. The histograms correspond to the distance between opposite ligands highlighted in red. Bottom: comparison between snapshots of ZIF-4-cp-II (left) and ZIF-4-cp (right) 6-membered rings. The histograms show the distribution of the swing dihedral angle formed by the atoms in red.}}
\end{figure}

We selected some microscopic features that allow to describe the changes observed in both phase transitions in terms of collective motions of the atoms in the unit cell. 
To do so, we performed individual simulations at thermodynamic conditions such that the two phases involved in each phase transition were stable separately, meaning that each of them is associated to a local minimum in the free energy surface.
For the ZIF-4/ZIF-4-cp transformation, the simulations were run at $P=20$ MPa and $T=300$ K. For both structures we monitored the distance between opposite ligands in the 8-membered Zn$^{2+}$ ring that constitutes the largest pore window. 
In the top panel of Fig. \ref{fig:histograms} we show snapshots of the two polymorphs with the mentioned distance marked in red. Histograms for this value in both phases are also plotted, showing a clear change that goes from $d\sim11$ \AA $\space$ in ZIF-4 to $d\sim8.5$ \AA $\space$ in ZIF-4-cp. These results are compatible with the previous characterization of the transition in terms of the closure of the unit cell pore\cite{Henke2018}. The values are also in good agreement with the experimental structures, that present distances of 11.8 \AA $\space$ for ZIF-4 and $8.5$ \AA $\space$ for ZIF-4-cp.

For studying the ZIF-4-cp/ZIF-4-cp-II transformation, we performed simulations at $P=140$ MPa and $T=300$ K. The microscopic changes undergone by the material within this transition are more subtle. 
Indeed, the unit cells of the two polymorphs do not show very clear differences in the Zn and N atomic positions, besides some slight distortions in the shape of the 6 and 8-membered rings. However, we noticed a change in the swing angles values of some of the ligands inside a 6-membered ring. In ZIF-4-cp, the tagged ligands have a C-H direction pointing to the external region of the cycle, while in ZIF-4-cp-II the ligand orientation is almost perpendicular to the rings plane (see Fig. \ref{fig:histograms}). This allows for a slightly better packing that contributes to the overall volume contraction. The swing angle in this kind of systems is usually defined as the dihedral angle between three consecutive Zn atoms and the central Carbon of the ligand that connects the last two of them. In the bottom part of Fig. \ref{fig:histograms} we show snapshots of both polymorphs, with the atoms considered for computing the swing angle marked in red. Despite the fact that the histograms for both phases superimpose, they are clearly centered in different values. The average angle for ZIF-4-cp is $\phi\sim-80^\circ$, while for ZIF-4-cp-II an average of $\phi\sim-105^\circ$ was computed.
In the experimental structures it is also possible to observe the swing angle change previously described, but the average angle for both species deviates from our histogram results. Values of $\phi=-106^\circ$ and $\phi=-179^\circ$ were computed from the \textit{cif} files of ZIF-4-cp and ZIF-4-cp-II respectively. These differences can be attributed to 
the high sensibility of the dihedral angle to small deviations in atom positions in cases where the four atoms are close to being aligned.

Our simulations suggest that pore closure and swing angle motions constitute the microscopic defects that propagate over the material to give rise to the phase transition. A full mechanistic study to determine for example whether the process is concerted or not,\cite{Wharmby2015} however, would imply considering larger simulation cells, as done for other MOFs by Rogge and coworkers.\cite{Rogge2019}

\subsection{ZIF-4 phase diagram}\label{phase}

To finalize the thermodynamic description of the ZIF-4 polymorphs, we aim to reconstruct the phase diagram. To do so, we extended the above phase equilibrium analyses to higher temperature regions via the Clapeyron relation, which allows us to obtain the slope of the coexistence curve $P(T)$ between two phases through the calculation of their enthalpy $H$ and volume differences. 

\begin{equation} \label{clapeyron}
    \frac{dP}{dT} = \frac{\Delta H}{T\Delta V} 
\end{equation}

For the calculation of these observables, it is enough to generate a stable trajectory of both phases separately at a certain point of the $(T,P)$ plane. This was done by performing unbiased simulations in the NPT ensemble starting from each one of the polymorphs. The values of $T$ and $P$ were chosen such that both phases were stable during the time span of the unbiased molecular dynamics simulations. In the case of ZIF-4-cp/ZIF-4-cp-II, a single long trajectory was enough to sample both states since a few interconvertion events took place. The resulting trajectory was split into regions corresponding to each polymorph to compute the desired observables. 
Once the slope of the coexistence curve is computed, the equilibrium pressures obtained in section \ref{stability} for ambient temperature can be extrapolated to other thermal conditions.
For condensed phases, it is reasonable to assume an approximately constant slope for all the coexistence interval.
In addition, we included in this analysis the glass phase obtained from thermal amorphization of ZIF-4: ZIF$\_$a\cite{Mendez2024}. 
For this transition, an equilibrium temperature in of $T=425$ K was calculated in a previous work\cite{Mendez2024} using the same force field at ambient pressure.
The thermodynamic information gathered from simulations of individual polymorphs is summarized in table \ref{tabla_termo}.

\begin{table}[h]
\caption{Thermodynamic information of the transitions considered for the construction of the phase diagram.}\label{tabla_termo}%
\begin{tabular}{@{}llllll@{}}
\toprule
               & $\Delta H$  & $\Delta V$&  $P$  & $T$ & $dP/dT$\\
               & (kJ/mol)  & (nm$^3$)&  (MPa)  & (K) & (MPa/K)\\
\midrule
ZIF-4 / ZIF-4-cp        & -12.4 & -0.52 & 20  & 425 & 0.093 \\
ZIF-4-cp / ZIF-4-cp-II  &  1.01 & -0.12 & 140 & 300 & -0.047 \\
ZIF-4 / ZIF\_a            & 55.4  & -0.66 & 0   & 425  &  -0.33 \\
ZIF-4-cp / ZIF\_a       & 66.0  & -0.075 & 20 & 425  & -3.44    \\
ZIF-4-cp-II / ZIF\_a      & 69.1 & 0.02  & 170  & 350 & 16.4 \\
\botrule
\end{tabular}
\end{table}

\begin{figure}[h]
\centering
\includegraphics[width=0.5\textwidth]{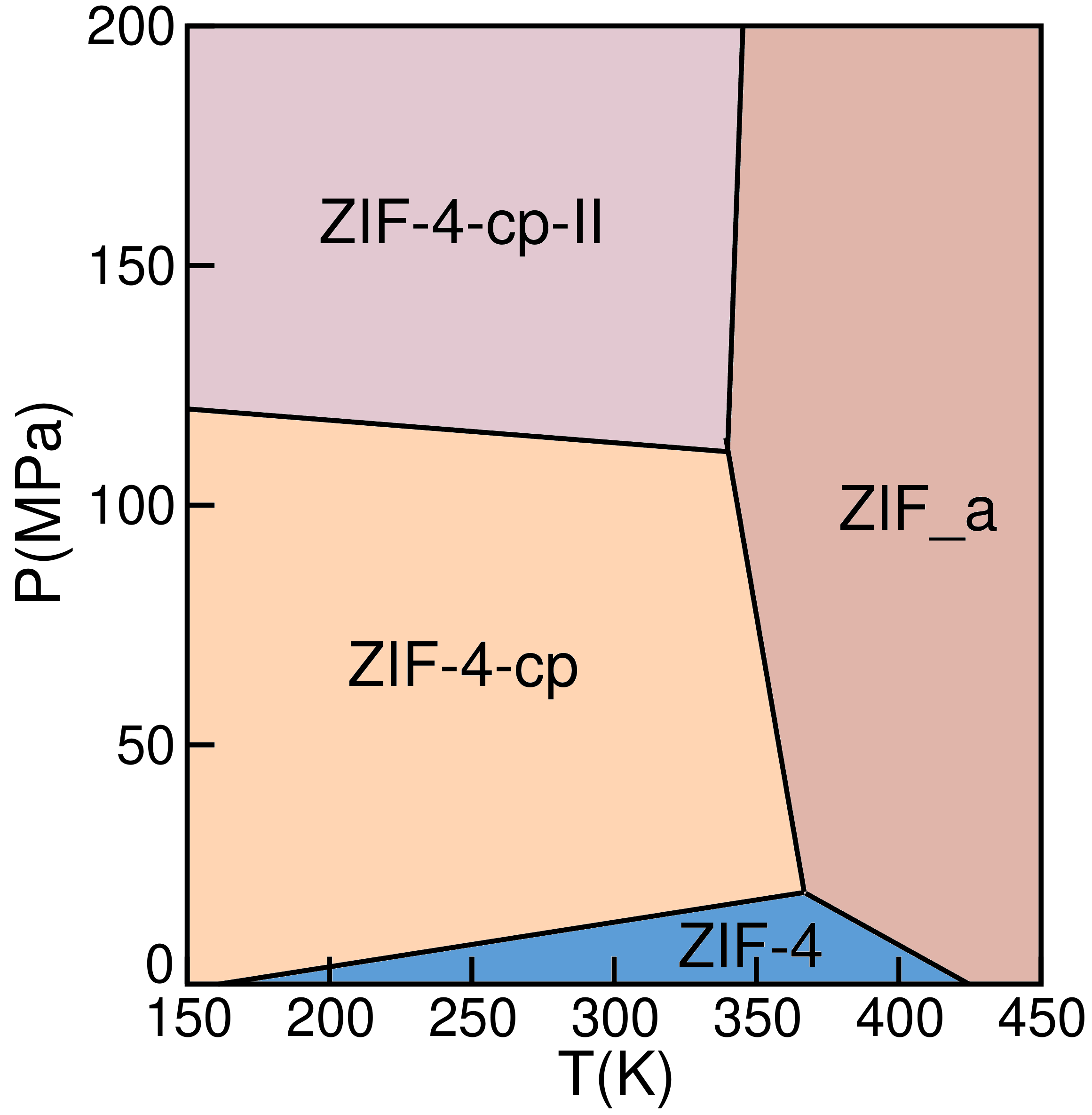}
\caption{\label{fig:phase_diagram}{Calculated ZIF-4 phase diagram in the T=150-450K and P=0-200 MPa region.}}
\end{figure}

The phase diagram was constructed using the information of the slopes of table \ref{tabla_termo}, together with the coexistence points at ambient temperature calculated in \ref{stability} and the one corresponding to ZIF-4/ZIF\_a, according to the following steps:
(i) The equilibrium points of ZIF-4/ZIF-4-cp and ZIF-4/ZIF\_a were extrapolated with the calculated slopes to find the triple point of these three species. 
(ii) From this triple point, the coexistence curve of ZIF-4-cp/ZIF\_a was extrapolated. 
(iii) The second triple point was obtained from the intersection of the curve from (ii) and the one obtained from the extrapolation of the equilibrium point of ZIF-4-cp/ZIF-4-cp-II at ambient temperature.
(iv) Finally, the coexistence curve between ZIF-4-cp-II and ZIF\_a was extrapolated from the triple point found in (iii).
The results are plotted in Fig. \ref{fig:phase_diagram}.

The computed phase diagram correctly reproduces the sign of the slope of ZIF-4/ZIF$\_$a and ZIF-4-cp-II/ZIF$\_$a equilibrium curves \cite{Widmer2019}. In both cases, the amorphous phase is higher in the enthalpy scale than the crystals. In terms of volume changes, the density of ZIF$\_$a lies in between the ZIF-4 and ZIF-4-cp-II values, thus giving rise to slopes of opposite sign as a final result. Our results match these tendencies.
ZIF-4-cp is not present in the reference phase diagram by Widmer and coworkers,\cite{Widmer2019} as explained in section \ref{stability}. Nevertheless, we can state that the slope of the ZIF-4/ZIF-4-cp is also correctly captured by our model, given the fact that this transition was also observed at ambient pressure and $T\sim140$ K\cite{Wharmby2015}. Our simulations predict this temperature to be around $T\sim160$ K. The positive sign of this slope means that the enthalpy of ZIF-4-cp must be lower than the one corresponding to ZIF-4, as our results suggest, since the change in volume is also negative.

\section{Conclusion}\label{sec13}

This work sheds light into the phase transformations that occur under pressure at temperatures surrounding ambient conditions for ZIF-4 and their associated polymorphs. We employ a classical force field which incorporates metal-ligand bonds reactivity within a well-tempered metadynamics scheme where the cell parameters are biased. These simulations allow to recover the ZIF-4, ZIF-4-cp and ZIF-4-cp-II phases and explore their stability regions. The multibaric formalism is applied to further explore a large range of thermodynamic conditions. The collective motions that drive the ZIF-4/ZIF-4-cp and ZIF-4-cp/ZIF-4-cp-II phase transformations are also described. The former one is related to pore closure while the latter one corresponds to a subtle change in swing dihedral angles. Our methodology is validated via the analysis of structural correlations between the experimental phases and the ones obtained in the well-tempered metadynamics simulations.      

Most importantly, we provide a sketch of the phase diagram of ZIF-4 and associated polymorphs in the T=150-450K, P=0-200 MPa range. This phase diagram has been elusive to experimental studies, since the large pressure ramps that have to be applied hamper the finding of all existing phases. Even though the absolute values of T,P in the critical points that we obtain do not match the absolute experimental values, the magnitude of the deviations are reasonable for classical force field-based simulations. Crucially, the most relevant qualitative tendencies, such as the sign of the phase separation curves, is correctly captured by our model, and the placement of the ZIF-4-cp phase in the phase diagram was achieved.  

Finally, we would like to highlight that our methodology can be extended to the treatment of other high-temperature/ high-pressure regions of the ZIF-4 phase diagram to answer remaining questions as for instance: what is the topology of the phase diagram surrounding ZIF-4-cp-III? Is there a phase transition between ZIF-4-cp-II and ZIF-hPT-II? Are ZIF-hPT-I and ZIF-hPT-II different polymorphs? These and more questions will be the object of further work.

\section*{Supplementary information}

Methodological details as described in the main text as well as sample input files and additional results supporting the convergence of the well-tempered metadynamics runs are available at the Supplementary Information.

\begin{acknowledgments}
This work was funded by the European Union ERC Starting
grant MAGNIFY, grant number 101042514. This work was
granted access to the HPC resources of CINES under the allocation
A0150911989 made by GENCI and to the HPC resources
of the MeSU platform at Sorbonne Université.
\end{acknowledgments}

\section*{Data Availability Statement}

The data that support the findings of this study are available within the article and its supplementary material.


\newpage

\MakeTitle{
        Supplementary Information of Phase Diagram of ZIF-4 Computed via Well-tempered Metadynamics
    }{
        Emilio Mendez and Rocio Semino           
    }{
    Sorbonne Université, CNRS, Physico-chimie des Electrolytes et Nanosystèmes Interfaciaux, PHENIX, F-75005 Paris, France
    }

\subsection{Well-tempered Metadynamics Setup}\label{sec0sup}

The following script was used as PLUMED input file for the configuration of the well-tempered metadynamics runs:

\begin{plumed}
# Define volume and cell parameters:
vol: VOLUME
cell: CELL

# Read Zn and N atom indexes from ndx file:
Zn_atoms: GROUP NDX_FILE=index.ndx NDX_GROUP=zn
N_atoms: GROUP NDX_FILE=index.ndx NDX_GROUP=n

# Calculate the Zn-N coordination number:
cn: COORDINATION GROUPA=Zn_atoms GROUPB=N_atoms R_0=0.05 D_0=0.22 NN=6

# Metadynamics setup:
meta: METAD ...
ARG=cell.ax,cell.by,cell.cz
SIGMA=0.1,0.1,0.1
HEIGHT=9.0
BIASFACTOR=75.0
TEMP=300.0
PACE=1000
GRID_MIN=2.4,2.4,2.4
GRID_MAX=4.4,4.4,4.4
GRID_BIN=200,200,200
 WALKERS_N=5
 WALKERS_ID=0
 WALKERS_DIR=../
 WALKERS_RSTRIDE=1000
...

# Additional constraints:
uwall: UPPER_WALLS ARG=vol,cell.ax,cell.by,cell.cz AT=36.0,3.8,3.8,3.8 KAPPA=10000,50000,50000,50000 EXP=2,2,2,2
lwall: LOWER_WALLS ARG=vol,cell.ax,cell.by,cell.cz,cn AT=16.0,2.6,2.6,2.6,511.0 KAPPA=10000,50000,50000,50000,50000 EXP=2,2,2,2,2

# Output:
PRINT ARG=vol,cell.ax,cell.by,cell.cz,uwall.bias,lwall.bias,meta.bias,cn FILE=colva.dat STRIDE=1000



\end{plumed}

Each section contains a comment explaining the aim of the code fragment that follows. Default PLUMED units were used: nm for distance, ps for time and kJ/mol for energy. The cell parameters $a$, $b$ and $c$ are called 'cell.ax', 'cell.by' and 'cell.cz'. The coordination number command 'COORDINATION' counts the number of Zn-N pairs at a distance lower than 0.22 nm, a slightly higher value than the typical bond length of 0.21 nm. This number is constrained  through the 'LOWER\_WALLS' command to be always higher than 511. Since the total coordination number in the system is 512, this prevents Zn-ligand bond breaking events. Additional descriptions of each command can be found in the PLUMED manual.

\subsection{Metadynamics Convergence and Data Treatment}\label{sec1sup}

\begin{figure}[h]
\centering
\includegraphics[width=0.7\textwidth]{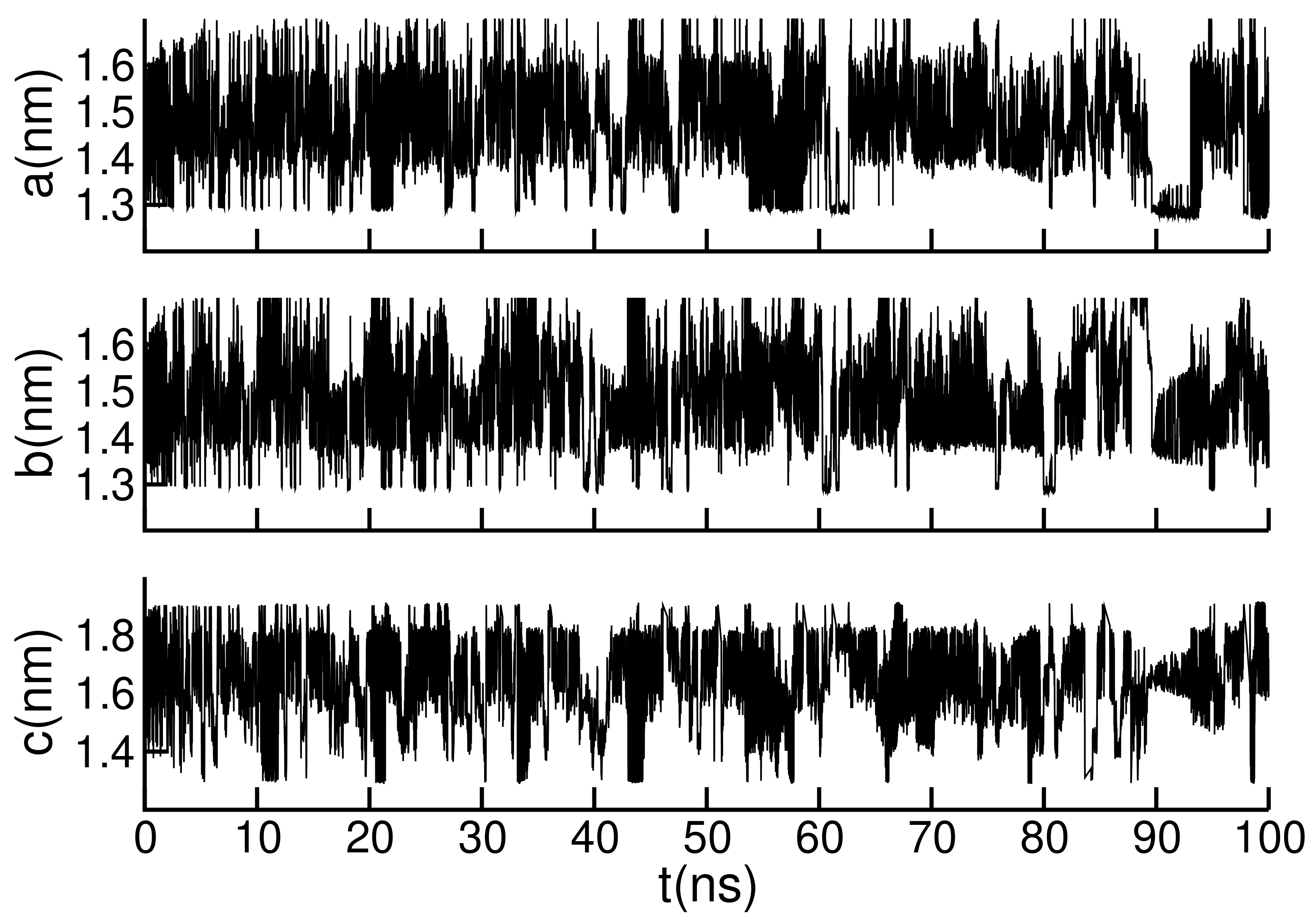}
\caption{\label{fig:cvs}{Collective variables $a$, $b$, and $c$ as a function of time for the $P=40$ MPa simulation.}}
\end{figure}

To check if the order parameter space was correctly sampled during the simulation, we plotted the evolution of $a$, $b$ and $c$ over time in Fig. S\ref{fig:cvs}, for the simulation at $P=40$ MPa. These results include the data from all the five walkers of the metadynamics.
Although it is not possible to directly assign each point to one of the studied polymorphs to observe the presence of interconvertion events, as it is usually done in metadynamics, it is clear that virtually all the points in the accessible region were visited multiple times, which is a necessary condition for the convergence of the method. Similar results were obtained for the simulations at $P=0$ and 80 MPa.

\begin{figure}[h]
\centering
\includegraphics[width=0.7\textwidth]{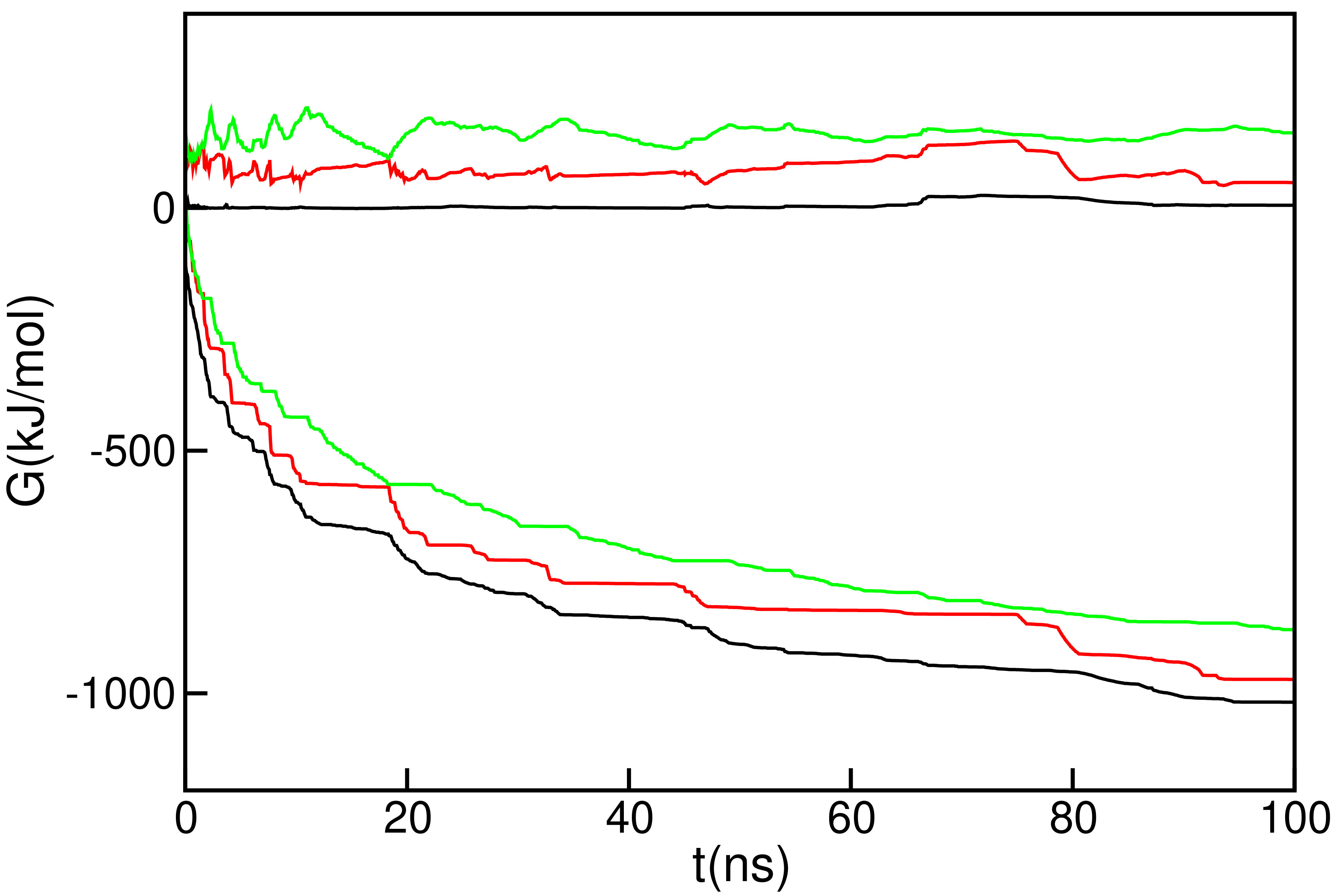}
\caption{\label{fig:estimator}{Free energy estimator for points in the $(a,b,c)$ space that corresponds to ZIF-4 (green), ZIF-4-cp-I (black) and ZIF-4-cp-II (red) lattice constants for the simulation at $P=40$ MPa. The curves in the negative region correspond to the estimator without the correction term of Eq. S\ref{free}.}}
\end{figure}

To analyze the convergence of the well-tempered metadynamics simulations we followed the procedure developed by Tiwary \textit{et al.} .\cite{Tiwary2014}
This approach takes into account the fact that in metadynamics the bias potential is dynamically modified as the simulation advances, never reaching a plateau value. This makes the choice of a convergence criterion a non trivial task.
The authors found a way to compute a time independent free energy estimator that allows to compare results measured at different times during the simulation given by:

\begin{equation} \label{free}
    G(s) = -\frac{\gamma V(s,t)}{(\gamma -1) } + k_b T \; \mathrm{ln} \int  e^{\frac{\gamma V(s,t)}{(\gamma -1) k_b T}} \; ds 
\end{equation}
were $s$ represents the collective variable(s), $\gamma$ the bias factor, and $V(s,t)$ the time dependent bias potential. 
The last term is a time dependent constant that aligns the free energy estimation at time $t$ with the ones computed at previous times. 
Equation S\ref{free} is only valid after the metadynamics has reached the convergence regime. 
For applying this technique to data obtained from different walkers, we time-ordered the gaussians coming from each simulation.
In Fig. S\ref{fig:estimator} we plotted the free energy estimator $G(a,b,c)$ of eq. \ref{free} as a function of time for three $(a,b,c)$ points that roughly correspond to the lattice constants of the studied polymorphs. We also plotted the free energy without the addition of the second term of eq. S\ref{free}.
As expected, these last values continue to descend without reaching a plateau, but the corrected estimators fluctuate around a fixed value after a transient period of time. 

\begin{figure}[h]
\centering
\includegraphics[width=0.7\textwidth]{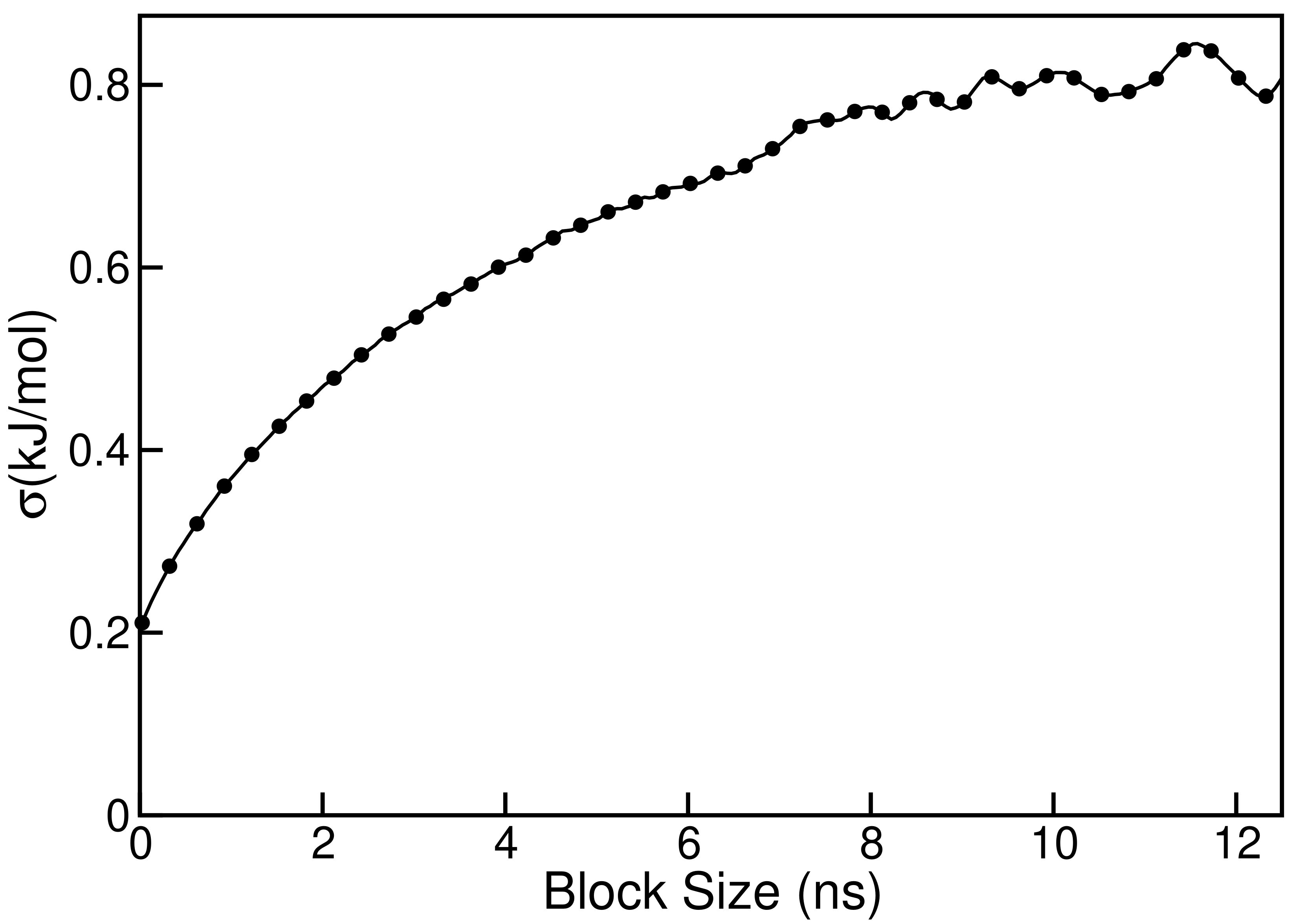}
\caption{\label{fig:block}{Free energy standard deviation for the absolute minimum as a function of the block size.}}
\end{figure}

In order to compute the final free energy profile and the corresponding errors, we need to time average the results from the corrected free energy curves. To avoid artifacts that arise when dealing with correlated data, we employed the block averaging technique developed by Bussi and Tribello.\cite{Bussi2019}
To estimate the optimal block size for which the data is uncorrelated, we computed the standard deviation of the free energy as a function of the block size. The results are shown in figure \ref{fig:block} for the $(a,b,c)$ point that corresponds to the lowest minimum. When the individual block values become uncorrelated, the standard deviation reaches a plateau. According to this criterion, we averaged data from blocks of 8 ns.

\subsection{Calculation of $G(V)$}\label{sec2sup}

For the computation of the free energy as a function of the volume at pressure $P$ ($G^{(P)}(V)$) starting from the function $G^{(P)}(a,b,c)$ that was obtained from the well-tempered metadynamics simulations, we proceeded as follows. First, we computed the probability $\mathcal{P}^{(P)}(a,b,c)$ given by the Boltzmann distribution:
\begin{equation} \label{boltz}
    \mathcal{P}^{(P)}(a,b,c) = C'e^{-\beta G^{(P)}(a,b,c)}
\end{equation}
where $C'$ is a normalization constant. Then, since the volume is an explicit function of $a$, $b$ and $c$, the probability of the system to have volume $V$ at a pressure $P$ was obtained by the following integration:
\begin{equation} \label{pv}
    \mathcal{P}^{(P)}(V) = \int \delta(V-abc) \; \mathcal{P}^{(P)}(a,b,c)\;da\;db\;dc
\end{equation}
where $\delta(x)$ is the Dirac delta function. In practice, this integral was performed by discretizing the volume variable into bins.
Finally $G(V)$ was obtained by the inversion of equation S\ref{boltz}, now with the volume as variable:
\begin{equation} \label{gv}
    G^{(P)}(V) = -1/\beta \; ln \; \mathcal{P}^{(P)}(V) 
\end{equation}
The corresponding error bars were calculated by propagation of errors from equations S\ref{boltz}, S\ref{pv} and S\ref{gv}.


\newpage

\bibliography{magnify}

\begin{thebibliography}{57}%
\makeatletter
\providecommand \@ifxundefined [1]{%
 \@ifx{#1\undefined}
}%
\providecommand \@ifnum [1]{%
 \ifnum #1\expandafter \@firstoftwo
 \else \expandafter \@secondoftwo
 \fi
}%
\providecommand \@ifx [1]{%
 \ifx #1\expandafter \@firstoftwo
 \else \expandafter \@secondoftwo
 \fi
}%
\providecommand \natexlab [1]{#1}%
\providecommand \enquote  [1]{``#1''}%
\providecommand \bibnamefont  [1]{#1}%
\providecommand \bibfnamefont [1]{#1}%
\providecommand \citenamefont [1]{#1}%
\providecommand \href@noop [0]{\@secondoftwo}%
\providecommand \href [0]{\begingroup \@sanitize@url \@href}%
\providecommand \@href[1]{\@@startlink{#1}\@@href}%
\providecommand \@@href[1]{\endgroup#1\@@endlink}%
\providecommand \@sanitize@url [0]{\catcode `\\12\catcode `\$12\catcode `\&12\catcode `\#12\catcode `\^12\catcode `\_12\catcode `\%12\relax}%
\providecommand \@@startlink[1]{}%
\providecommand \@@endlink[0]{}%
\providecommand \url  [0]{\begingroup\@sanitize@url \@url }%
\providecommand \@url [1]{\endgroup\@href {#1}{\urlprefix }}%
\providecommand \urlprefix  [0]{URL }%
\providecommand \Eprint [0]{\href }%
\providecommand \doibase [0]{http://dx.doi.org/}%
\providecommand \selectlanguage [0]{\@gobble}%
\providecommand \bibinfo  [0]{\@secondoftwo}%
\providecommand \bibfield  [0]{\@secondoftwo}%
\providecommand \translation [1]{[#1]}%
\providecommand \BibitemOpen [0]{}%
\providecommand \bibitemStop [0]{}%
\providecommand \bibitemNoStop [0]{.\EOS\space}%
\providecommand \EOS [0]{\spacefactor3000\relax}%
\providecommand \BibitemShut  [1]{\csname bibitem#1\endcsname}%
\let\auto@bib@innerbib\@empty
\bibitem [{\citenamefont {Park}\ \emph {et~al.}(2006)\citenamefont {Park}, \citenamefont {Ni}, \citenamefont {C{\^{o}}t{\'{e}}}, \citenamefont {Choi}, \citenamefont {Huang}, \citenamefont {Uribe-Romo}, \citenamefont {Chae}, \citenamefont {O'Keeffe},\ and\ \citenamefont {Yaghi}}]{Park2006}%
  \BibitemOpen
  \bibfield  {author} {\bibinfo {author} {\bibfnamefont {K.~S.}\ \bibnamefont {Park}}, \bibinfo {author} {\bibfnamefont {Z.}~\bibnamefont {Ni}}, \bibinfo {author} {\bibfnamefont {A.~P.}\ \bibnamefont {C{\^{o}}t{\'{e}}}}, \bibinfo {author} {\bibfnamefont {J.~Y.}\ \bibnamefont {Choi}}, \bibinfo {author} {\bibfnamefont {R.}~\bibnamefont {Huang}}, \bibinfo {author} {\bibfnamefont {F.~J.}\ \bibnamefont {Uribe-Romo}}, \bibinfo {author} {\bibfnamefont {H.~K.}\ \bibnamefont {Chae}}, \bibinfo {author} {\bibfnamefont {M.}~\bibnamefont {O'Keeffe}}, \ and\ \bibinfo {author} {\bibfnamefont {O.~M.}\ \bibnamefont {Yaghi}},\ }\bibfield  {title} {\enquote {\bibinfo {title} {Exceptional chemical and thermal stability of zeolitic imidazolate frameworks},}\ }\href {\doibase 10.1073/pnas.0602439103} {\bibfield  {journal} {\bibinfo  {journal} {Proceedings of the National Academy of Sciences}\ }\textbf {\bibinfo {volume} {103}},\ \bibinfo {pages} {10186--10191} (\bibinfo {year} {2006})}\BibitemShut {NoStop}%
\bibitem [{\citenamefont {Widmer}\ \emph {et~al.}(2019{\natexlab{a}})\citenamefont {Widmer}, \citenamefont {Lampronti}, \citenamefont {Chibani}, \citenamefont {Wilson}, \citenamefont {Anzellini}, \citenamefont {Farsang}, \citenamefont {Kleppe}, \citenamefont {Casati}, \citenamefont {MacLeod}, \citenamefont {Redfern}, \citenamefont {Coudert},\ and\ \citenamefont {Bennett}}]{Widmer2019}%
  \BibitemOpen
  \bibfield  {author} {\bibinfo {author} {\bibfnamefont {R.~N.}\ \bibnamefont {Widmer}}, \bibinfo {author} {\bibfnamefont {G.~I.}\ \bibnamefont {Lampronti}}, \bibinfo {author} {\bibfnamefont {S.}~\bibnamefont {Chibani}}, \bibinfo {author} {\bibfnamefont {C.~W.}\ \bibnamefont {Wilson}}, \bibinfo {author} {\bibfnamefont {S.}~\bibnamefont {Anzellini}}, \bibinfo {author} {\bibfnamefont {S.}~\bibnamefont {Farsang}}, \bibinfo {author} {\bibfnamefont {A.~K.}\ \bibnamefont {Kleppe}}, \bibinfo {author} {\bibfnamefont {N.~P.~M.}\ \bibnamefont {Casati}}, \bibinfo {author} {\bibfnamefont {S.~G.}\ \bibnamefont {MacLeod}}, \bibinfo {author} {\bibfnamefont {S.~A.~T.}\ \bibnamefont {Redfern}}, \bibinfo {author} {\bibfnamefont {F.-X.}\ \bibnamefont {Coudert}}, \ and\ \bibinfo {author} {\bibfnamefont {T.~D.}\ \bibnamefont {Bennett}},\ }\bibfield  {title} {\enquote {\bibinfo {title} {Rich polymorphism of a metal{\textendash}organic framework in pressure{\textendash}temperature space},}\ }\href {\doibase 10.1021/jacs.9b03234}
  {\bibfield  {journal} {\bibinfo  {journal} {Journal of the American Chemical Society}\ }\textbf {\bibinfo {volume} {141}},\ \bibinfo {pages} {9330--9337} (\bibinfo {year} {2019}{\natexlab{a}})}\BibitemShut {NoStop}%
\bibitem [{\citenamefont {Bennett}\ \emph {et~al.}(2010)\citenamefont {Bennett}, \citenamefont {Goodwin}, \citenamefont {Dove}, \citenamefont {Keen}, \citenamefont {Tucker}, \citenamefont {Barney}, \citenamefont {Soper}, \citenamefont {Bithell}, \citenamefont {Tan},\ and\ \citenamefont {Cheetham}}]{Bennett2010}%
  \BibitemOpen
  \bibfield  {author} {\bibinfo {author} {\bibfnamefont {T.~D.}\ \bibnamefont {Bennett}}, \bibinfo {author} {\bibfnamefont {A.~L.}\ \bibnamefont {Goodwin}}, \bibinfo {author} {\bibfnamefont {M.~T.}\ \bibnamefont {Dove}}, \bibinfo {author} {\bibfnamefont {D.~A.}\ \bibnamefont {Keen}}, \bibinfo {author} {\bibfnamefont {M.~G.}\ \bibnamefont {Tucker}}, \bibinfo {author} {\bibfnamefont {E.~R.}\ \bibnamefont {Barney}}, \bibinfo {author} {\bibfnamefont {A.~K.}\ \bibnamefont {Soper}}, \bibinfo {author} {\bibfnamefont {E.~G.}\ \bibnamefont {Bithell}}, \bibinfo {author} {\bibfnamefont {J.-C.}\ \bibnamefont {Tan}}, \ and\ \bibinfo {author} {\bibfnamefont {A.~K.}\ \bibnamefont {Cheetham}},\ }\bibfield  {title} {\enquote {\bibinfo {title} {Structure and properties of an amorphous metal-organic framework},}\ }\href {\doibase 10.1103/physrevlett.104.115503} {\bibfield  {journal} {\bibinfo  {journal} {Physical Review Letters}\ }\textbf {\bibinfo {volume} {104}},\ \bibinfo {pages} {115503} (\bibinfo {year}
  {2010})}\BibitemShut {NoStop}%
\bibitem [{\citenamefont {Gaillac}\ \emph {et~al.}(2017)\citenamefont {Gaillac}, \citenamefont {Pullumbi}, \citenamefont {Beyer}, \citenamefont {Chapman}, \citenamefont {Keen}, \citenamefont {Bennett},\ and\ \citenamefont {Coudert}}]{Gaillac2017}%
  \BibitemOpen
  \bibfield  {author} {\bibinfo {author} {\bibfnamefont {R.}~\bibnamefont {Gaillac}}, \bibinfo {author} {\bibfnamefont {P.}~\bibnamefont {Pullumbi}}, \bibinfo {author} {\bibfnamefont {K.~A.}\ \bibnamefont {Beyer}}, \bibinfo {author} {\bibfnamefont {K.~W.}\ \bibnamefont {Chapman}}, \bibinfo {author} {\bibfnamefont {D.~A.}\ \bibnamefont {Keen}}, \bibinfo {author} {\bibfnamefont {T.~D.}\ \bibnamefont {Bennett}}, \ and\ \bibinfo {author} {\bibfnamefont {F.-X.}\ \bibnamefont {Coudert}},\ }\bibfield  {title} {\enquote {\bibinfo {title} {Liquid metal{\textendash}organic frameworks},}\ }\href {\doibase 10.1038/nmat4998} {\bibfield  {journal} {\bibinfo  {journal} {Nature Materials}\ }\textbf {\bibinfo {volume} {16}},\ \bibinfo {pages} {1149--1154} (\bibinfo {year} {2017})}\BibitemShut {NoStop}%
\bibitem [{\citenamefont {Bennett}\ \emph {et~al.}(2015)\citenamefont {Bennett}, \citenamefont {Tan}, \citenamefont {Yue}, \citenamefont {Baxter}, \citenamefont {Ducati}, \citenamefont {Terrill}, \citenamefont {Yeung}, \citenamefont {Zhou}, \citenamefont {Chen}, \citenamefont {Henke}, \citenamefont {Cheetham},\ and\ \citenamefont {Greaves}}]{Bennett2015}%
  \BibitemOpen
  \bibfield  {author} {\bibinfo {author} {\bibfnamefont {T.~D.}\ \bibnamefont {Bennett}}, \bibinfo {author} {\bibfnamefont {J.-C.}\ \bibnamefont {Tan}}, \bibinfo {author} {\bibfnamefont {Y.}~\bibnamefont {Yue}}, \bibinfo {author} {\bibfnamefont {E.}~\bibnamefont {Baxter}}, \bibinfo {author} {\bibfnamefont {C.}~\bibnamefont {Ducati}}, \bibinfo {author} {\bibfnamefont {N.~J.}\ \bibnamefont {Terrill}}, \bibinfo {author} {\bibfnamefont {H.~H.~M.}\ \bibnamefont {Yeung}}, \bibinfo {author} {\bibfnamefont {Z.}~\bibnamefont {Zhou}}, \bibinfo {author} {\bibfnamefont {W.}~\bibnamefont {Chen}}, \bibinfo {author} {\bibfnamefont {S.}~\bibnamefont {Henke}}, \bibinfo {author} {\bibfnamefont {A.~K.}\ \bibnamefont {Cheetham}}, \ and\ \bibinfo {author} {\bibfnamefont {G.~N.}\ \bibnamefont {Greaves}},\ }\bibfield  {title} {\enquote {\bibinfo {title} {Hybrid glasses from strong and fragile metal-organic framework liquids},}\ }\href {\doibase 10.1038/ncomms9079} {\bibfield  {journal} {\bibinfo  {journal} {Nature Communications}\
  }\textbf {\bibinfo {volume} {6}},\ \bibinfo {pages} {8079} (\bibinfo {year} {2015})}\BibitemShut {NoStop}%
\bibitem [{\citenamefont {Bennett}\ and\ \citenamefont {Horike}(2018)}]{Bennett2018}%
  \BibitemOpen
  \bibfield  {author} {\bibinfo {author} {\bibfnamefont {T.~D.}\ \bibnamefont {Bennett}}\ and\ \bibinfo {author} {\bibfnamefont {S.}~\bibnamefont {Horike}},\ }\bibfield  {title} {\enquote {\bibinfo {title} {Liquid, glass and amorphous solid states of coordination polymers and metal–organic frameworks},}\ }\href {\doibase 10.1038/s41578-018-0054-3} {\bibfield  {journal} {\bibinfo  {journal} {Nature Reviews Materials}\ }\textbf {\bibinfo {volume} {3}},\ \bibinfo {pages} {431–440} (\bibinfo {year} {2018})}\BibitemShut {NoStop}%
\bibitem [{\citenamefont {Madsen}\ \emph {et~al.}(2020)\citenamefont {Madsen}, \citenamefont {Qiao}, \citenamefont {Sen}, \citenamefont {Hung}, \citenamefont {Chen}, \citenamefont {Gan}, \citenamefont {Sen},\ and\ \citenamefont {Yue}}]{Madsen2020}%
  \BibitemOpen
  \bibfield  {author} {\bibinfo {author} {\bibfnamefont {R.~S.~K.}\ \bibnamefont {Madsen}}, \bibinfo {author} {\bibfnamefont {A.}~\bibnamefont {Qiao}}, \bibinfo {author} {\bibfnamefont {J.}~\bibnamefont {Sen}}, \bibinfo {author} {\bibfnamefont {I.}~\bibnamefont {Hung}}, \bibinfo {author} {\bibfnamefont {K.}~\bibnamefont {Chen}}, \bibinfo {author} {\bibfnamefont {Z.}~\bibnamefont {Gan}}, \bibinfo {author} {\bibfnamefont {S.}~\bibnamefont {Sen}}, \ and\ \bibinfo {author} {\bibfnamefont {Y.}~\bibnamefont {Yue}},\ }\bibfield  {title} {\enquote {\bibinfo {title} {Ultrahigh-field 67 zn nmr reveals short-range disorder in zeolitic imidazolate framework glasses},}\ }\href {\doibase 10.1126/science.aaz0251} {\bibfield  {journal} {\bibinfo  {journal} {Science}\ }\textbf {\bibinfo {volume} {367}},\ \bibinfo {pages} {1473–1476} (\bibinfo {year} {2020})}\BibitemShut {NoStop}%
\bibitem [{\citenamefont {Weiß}\ and\ \citenamefont {Henke}(2023)}]{Wei2023}%
  \BibitemOpen
  \bibfield  {author} {\bibinfo {author} {\bibfnamefont {J.-B.}\ \bibnamefont {Weiß}}\ and\ \bibinfo {author} {\bibfnamefont {S.}~\bibnamefont {Henke}},\ }\bibfield  {title} {\enquote {\bibinfo {title} {Forging links in molecular glasses},}\ }\href {\doibase 10.1038/s44160-023-00425-0} {\bibfield  {journal} {\bibinfo  {journal} {Nature Synthesis}\ }\textbf {\bibinfo {volume} {3}},\ \bibinfo {pages} {150–151} (\bibinfo {year} {2023})}\BibitemShut {NoStop}%
\bibitem [{\citenamefont {Kim}\ \emph {et~al.}(2024)\citenamefont {Kim}, \citenamefont {Lee}, \citenamefont {Seo}, \citenamefont {Cho}, \citenamefont {Jeon},\ and\ \citenamefont {Moon}}]{Kim2024}%
  \BibitemOpen
  \bibfield  {author} {\bibinfo {author} {\bibfnamefont {M.}~\bibnamefont {Kim}}, \bibinfo {author} {\bibfnamefont {H.-S.}\ \bibnamefont {Lee}}, \bibinfo {author} {\bibfnamefont {D.-H.}\ \bibnamefont {Seo}}, \bibinfo {author} {\bibfnamefont {S.~J.}\ \bibnamefont {Cho}}, \bibinfo {author} {\bibfnamefont {E.-c.}\ \bibnamefont {Jeon}}, \ and\ \bibinfo {author} {\bibfnamefont {H.~R.}\ \bibnamefont {Moon}},\ }\bibfield  {title} {\enquote {\bibinfo {title} {Melt-quenched carboxylate metal–organic framework glasses},}\ }\href {\doibase 10.1038/s41467-024-45326-8} {\bibfield  {journal} {\bibinfo  {journal} {Nature Communications}\ }\textbf {\bibinfo {volume} {15}} (\bibinfo {year} {2024}),\ 10.1038/s41467-024-45326-8}\BibitemShut {NoStop}%
\bibitem [{\citenamefont {Xue}\ \emph {et~al.}(2024)\citenamefont {Xue}, \citenamefont {Das}, \citenamefont {Weiß},\ and\ \citenamefont {Henke}}]{Xue2024}%
  \BibitemOpen
  \bibfield  {author} {\bibinfo {author} {\bibfnamefont {W.-L.}\ \bibnamefont {Xue}}, \bibinfo {author} {\bibfnamefont {C.}~\bibnamefont {Das}}, \bibinfo {author} {\bibfnamefont {J.-B.}\ \bibnamefont {Weiß}}, \ and\ \bibinfo {author} {\bibfnamefont {S.}~\bibnamefont {Henke}},\ }\bibfield  {title} {\enquote {\bibinfo {title} {Insights into the mechanochemical glass formation of zeolitic imidazolate frameworks},}\ }\href {\doibase 10.1002/anie.202405307} {\bibfield  {journal} {\bibinfo  {journal} {Angewandte Chemie International Edition}\ } (\bibinfo {year} {2024}),\ 10.1002/anie.202405307}\BibitemShut {NoStop}%
\bibitem [{\citenamefont {Hartmann}\ \emph {et~al.}(2015)\citenamefont {Hartmann}, \citenamefont {B\"{o}hme}, \citenamefont {Hovestadt},\ and\ \citenamefont {Paula}}]{Hartmann2015}%
  \BibitemOpen
  \bibfield  {author} {\bibinfo {author} {\bibfnamefont {M.}~\bibnamefont {Hartmann}}, \bibinfo {author} {\bibfnamefont {U.}~\bibnamefont {B\"{o}hme}}, \bibinfo {author} {\bibfnamefont {M.}~\bibnamefont {Hovestadt}}, \ and\ \bibinfo {author} {\bibfnamefont {C.}~\bibnamefont {Paula}},\ }\bibfield  {title} {\enquote {\bibinfo {title} {Adsorptive separation of olefin/paraffin mixtures with {ZIF}-4},}\ }\href {\doibase 10.1021/acs.langmuir.5b02907} {\bibfield  {journal} {\bibinfo  {journal} {Langmuir}\ }\textbf {\bibinfo {volume} {31}},\ \bibinfo {pages} {12382--12389} (\bibinfo {year} {2015})}\BibitemShut {NoStop}%
\bibitem [{\citenamefont {Hovestadt}\ \emph {et~al.}(2018)\citenamefont {Hovestadt}, \citenamefont {Friebe}, \citenamefont {Helmich}, \citenamefont {Lange}, \citenamefont {M\"{o}llmer}, \citenamefont {Gl\"{a}ser}, \citenamefont {Mundstock},\ and\ \citenamefont {Hartmann}}]{Hovestadt2018}%
  \BibitemOpen
  \bibfield  {author} {\bibinfo {author} {\bibfnamefont {M.}~\bibnamefont {Hovestadt}}, \bibinfo {author} {\bibfnamefont {S.}~\bibnamefont {Friebe}}, \bibinfo {author} {\bibfnamefont {L.}~\bibnamefont {Helmich}}, \bibinfo {author} {\bibfnamefont {M.}~\bibnamefont {Lange}}, \bibinfo {author} {\bibfnamefont {J.}~\bibnamefont {M\"{o}llmer}}, \bibinfo {author} {\bibfnamefont {R.}~\bibnamefont {Gl\"{a}ser}}, \bibinfo {author} {\bibfnamefont {A.}~\bibnamefont {Mundstock}}, \ and\ \bibinfo {author} {\bibfnamefont {M.}~\bibnamefont {Hartmann}},\ }\bibfield  {title} {\enquote {\bibinfo {title} {Continuous separation of light olefin/paraffin mixtures on zif-4 by pressure swing adsorption and membrane permeation},}\ }\href {\doibase 10.3390/molecules23040889} {\bibfield  {journal} {\bibinfo  {journal} {Molecules}\ }\textbf {\bibinfo {volume} {23}},\ \bibinfo {pages} {889} (\bibinfo {year} {2018})}\BibitemShut {NoStop}%
\bibitem [{\citenamefont {Bazer-Bachi}\ \emph {et~al.}(2014)\citenamefont {Bazer-Bachi}, \citenamefont {Assi{\'{e}}}, \citenamefont {Lecocq}, \citenamefont {Harbuzaru},\ and\ \citenamefont {Falk}}]{BazerBachi2014}%
  \BibitemOpen
  \bibfield  {author} {\bibinfo {author} {\bibfnamefont {D.}~\bibnamefont {Bazer-Bachi}}, \bibinfo {author} {\bibfnamefont {L.}~\bibnamefont {Assi{\'{e}}}}, \bibinfo {author} {\bibfnamefont {V.}~\bibnamefont {Lecocq}}, \bibinfo {author} {\bibfnamefont {B.}~\bibnamefont {Harbuzaru}}, \ and\ \bibinfo {author} {\bibfnamefont {V.}~\bibnamefont {Falk}},\ }\bibfield  {title} {\enquote {\bibinfo {title} {Towards industrial use of metal-organic framework: Impact of shaping on the {MOF} properties},}\ }\href {\doibase 10.1016/j.powtec.2013.09.013} {\bibfield  {journal} {\bibinfo  {journal} {Powder Technology}\ }\textbf {\bibinfo {volume} {255}},\ \bibinfo {pages} {52--59} (\bibinfo {year} {2014})}\BibitemShut {NoStop}%
\bibitem [{\citenamefont {Hovestadt}\ \emph {et~al.}(2017)\citenamefont {Hovestadt}, \citenamefont {Schmitz}, \citenamefont {Weissenberger}, \citenamefont {Reif}, \citenamefont {Kaspereit}, \citenamefont {Schwieger},\ and\ \citenamefont {Hartmann}}]{Hovestadt2017}%
  \BibitemOpen
  \bibfield  {author} {\bibinfo {author} {\bibfnamefont {M.}~\bibnamefont {Hovestadt}}, \bibinfo {author} {\bibfnamefont {J.~V.}\ \bibnamefont {Schmitz}}, \bibinfo {author} {\bibfnamefont {T.}~\bibnamefont {Weissenberger}}, \bibinfo {author} {\bibfnamefont {F.}~\bibnamefont {Reif}}, \bibinfo {author} {\bibfnamefont {M.}~\bibnamefont {Kaspereit}}, \bibinfo {author} {\bibfnamefont {W.}~\bibnamefont {Schwieger}}, \ and\ \bibinfo {author} {\bibfnamefont {M.}~\bibnamefont {Hartmann}},\ }\bibfield  {title} {\enquote {\bibinfo {title} {Scale-up of the synthesis of zeolitic imidazolate framework {ZIF}-4},}\ }\href {\doibase 10.1002/cite.201700105} {\bibfield  {journal} {\bibinfo  {journal} {Chemie Ingenieur Technik}\ }\textbf {\bibinfo {volume} {89}},\ \bibinfo {pages} {1374--1378} (\bibinfo {year} {2017})}\BibitemShut {NoStop}%
\bibitem [{\citenamefont {Frentzel-Beyme}\ \emph {et~al.}(2019)\citenamefont {Frentzel-Beyme}, \citenamefont {Kloß}, \citenamefont {Kolodzeiski}, \citenamefont {Pallach},\ and\ \citenamefont {Henke}}]{FrentzelBeyme2019}%
  \BibitemOpen
  \bibfield  {author} {\bibinfo {author} {\bibfnamefont {L.}~\bibnamefont {Frentzel-Beyme}}, \bibinfo {author} {\bibfnamefont {M.}~\bibnamefont {Kloß}}, \bibinfo {author} {\bibfnamefont {P.}~\bibnamefont {Kolodzeiski}}, \bibinfo {author} {\bibfnamefont {R.}~\bibnamefont {Pallach}}, \ and\ \bibinfo {author} {\bibfnamefont {S.}~\bibnamefont {Henke}},\ }\bibfield  {title} {\enquote {\bibinfo {title} {Meltable mixed-linker zeolitic imidazolate frameworks and their microporous glasses: From melting point engineering to selective hydrocarbon sorption},}\ }\href {\doibase 10.1021/jacs.9b05558} {\bibfield  {journal} {\bibinfo  {journal} {Journal of the American Chemical Society}\ }\textbf {\bibinfo {volume} {141}},\ \bibinfo {pages} {12362–12371} (\bibinfo {year} {2019})}\BibitemShut {NoStop}%
\bibitem [{\citenamefont {Yu}\ \emph {et~al.}(2020)\citenamefont {Yu}, \citenamefont {Qiao}, \citenamefont {Bumstead}, \citenamefont {Bennett}, \citenamefont {Yue},\ and\ \citenamefont {Tao}}]{Yu2020}%
  \BibitemOpen
  \bibfield  {author} {\bibinfo {author} {\bibfnamefont {Y.}~\bibnamefont {Yu}}, \bibinfo {author} {\bibfnamefont {A.}~\bibnamefont {Qiao}}, \bibinfo {author} {\bibfnamefont {A.~M.}\ \bibnamefont {Bumstead}}, \bibinfo {author} {\bibfnamefont {T.~D.}\ \bibnamefont {Bennett}}, \bibinfo {author} {\bibfnamefont {Y.}~\bibnamefont {Yue}}, \ and\ \bibinfo {author} {\bibfnamefont {H.}~\bibnamefont {Tao}},\ }\bibfield  {title} {\enquote {\bibinfo {title} {Impact of 1-methylimidazole on crystal formation, phase transitions, and glass formation in a zeolitic imidazolate framework},}\ }\href {\doibase 10.1021/acs.cgd.0c00740} {\bibfield  {journal} {\bibinfo  {journal} {Crystal Growth \& Design}\ }\textbf {\bibinfo {volume} {20}},\ \bibinfo {pages} {6528–6534} (\bibinfo {year} {2020})}\BibitemShut {NoStop}%
\bibitem [{\citenamefont {Iacomi}\ and\ \citenamefont {Maurin}(2021)}]{Iacomi2021}%
  \BibitemOpen
  \bibfield  {author} {\bibinfo {author} {\bibfnamefont {P.}~\bibnamefont {Iacomi}}\ and\ \bibinfo {author} {\bibfnamefont {G.}~\bibnamefont {Maurin}},\ }\bibfield  {title} {\enquote {\bibinfo {title} {Responzif structures: Zeolitic imidazolate frameworks as stimuli-responsive materials},}\ }\href {\doibase 10.1021/acsami.1c12403} {\bibfield  {journal} {\bibinfo  {journal} {ACS Applied Materials \& Interfaces}\ }\textbf {\bibinfo {volume} {13}},\ \bibinfo {pages} {50602–50642} (\bibinfo {year} {2021})}\BibitemShut {NoStop}%
\bibitem [{\citenamefont {Bennett}\ \emph {et~al.}(2011{\natexlab{a}})\citenamefont {Bennett}, \citenamefont {Keen}, \citenamefont {Tan}, \citenamefont {Barney}, \citenamefont {Goodwin},\ and\ \citenamefont {Cheetham}}]{Bennett2011}%
  \BibitemOpen
  \bibfield  {author} {\bibinfo {author} {\bibfnamefont {T.~D.}\ \bibnamefont {Bennett}}, \bibinfo {author} {\bibfnamefont {D.~A.}\ \bibnamefont {Keen}}, \bibinfo {author} {\bibfnamefont {J.-C.}\ \bibnamefont {Tan}}, \bibinfo {author} {\bibfnamefont {E.~R.}\ \bibnamefont {Barney}}, \bibinfo {author} {\bibfnamefont {A.~L.}\ \bibnamefont {Goodwin}}, \ and\ \bibinfo {author} {\bibfnamefont {A.~K.}\ \bibnamefont {Cheetham}},\ }\bibfield  {title} {\enquote {\bibinfo {title} {Thermal amorphization of zeolitic imidazolate frameworks},}\ }\href {\doibase 10.1002/anie.201007303} {\bibfield  {journal} {\bibinfo  {journal} {Angewandte Chemie International Edition}\ }\textbf {\bibinfo {volume} {50}},\ \bibinfo {pages} {3067--3071} (\bibinfo {year} {2011}{\natexlab{a}})}\BibitemShut {NoStop}%
\bibitem [{\citenamefont {Bennett}\ \emph {et~al.}(2011{\natexlab{b}})\citenamefont {Bennett}, \citenamefont {Simoncic}, \citenamefont {Moggach}, \citenamefont {Gozzo}, \citenamefont {Macchi}, \citenamefont {Keen}, \citenamefont {Tan},\ and\ \citenamefont {Cheetham}}]{Bennett2011_3}%
  \BibitemOpen
  \bibfield  {author} {\bibinfo {author} {\bibfnamefont {T.~D.}\ \bibnamefont {Bennett}}, \bibinfo {author} {\bibfnamefont {P.}~\bibnamefont {Simoncic}}, \bibinfo {author} {\bibfnamefont {S.~A.}\ \bibnamefont {Moggach}}, \bibinfo {author} {\bibfnamefont {F.}~\bibnamefont {Gozzo}}, \bibinfo {author} {\bibfnamefont {P.}~\bibnamefont {Macchi}}, \bibinfo {author} {\bibfnamefont {D.~A.}\ \bibnamefont {Keen}}, \bibinfo {author} {\bibfnamefont {J.-C.}\ \bibnamefont {Tan}}, \ and\ \bibinfo {author} {\bibfnamefont {A.~K.}\ \bibnamefont {Cheetham}},\ }\bibfield  {title} {\enquote {\bibinfo {title} {Reversible pressure-induced amorphization of a zeolitic imidazolate framework ({ZIF}-4)},}\ }\href {\doibase 10.1039/c1cc11985k} {\bibfield  {journal} {\bibinfo  {journal} {Chemical Communications}\ }\textbf {\bibinfo {volume} {47}},\ \bibinfo {pages} {7983} (\bibinfo {year} {2011}{\natexlab{b}})}\BibitemShut {NoStop}%
\bibitem [{\citenamefont {Henke}\ \emph {et~al.}(2018)\citenamefont {Henke}, \citenamefont {Wharmby}, \citenamefont {Kieslich}, \citenamefont {Hante}, \citenamefont {Schneemann}, \citenamefont {Wu}, \citenamefont {Daisenberger},\ and\ \citenamefont {Cheetham}}]{Henke2018}%
  \BibitemOpen
  \bibfield  {author} {\bibinfo {author} {\bibfnamefont {S.}~\bibnamefont {Henke}}, \bibinfo {author} {\bibfnamefont {M.~T.}\ \bibnamefont {Wharmby}}, \bibinfo {author} {\bibfnamefont {G.}~\bibnamefont {Kieslich}}, \bibinfo {author} {\bibfnamefont {I.}~\bibnamefont {Hante}}, \bibinfo {author} {\bibfnamefont {A.}~\bibnamefont {Schneemann}}, \bibinfo {author} {\bibfnamefont {Y.}~\bibnamefont {Wu}}, \bibinfo {author} {\bibfnamefont {D.}~\bibnamefont {Daisenberger}}, \ and\ \bibinfo {author} {\bibfnamefont {A.~K.}\ \bibnamefont {Cheetham}},\ }\bibfield  {title} {\enquote {\bibinfo {title} {Pore closure in zeolitic imidazolate frameworks under mechanical pressure},}\ }\href {\doibase 10.1039/c7sc04952h} {\bibfield  {journal} {\bibinfo  {journal} {Chemical Science}\ }\textbf {\bibinfo {volume} {9}},\ \bibinfo {pages} {1654–1660} (\bibinfo {year} {2018})}\BibitemShut {NoStop}%
\bibitem [{\citenamefont {Vervoorts}\ \emph {et~al.}(2019)\citenamefont {Vervoorts}, \citenamefont {Hobday}, \citenamefont {Ehrenreich}, \citenamefont {Daisenberger},\ and\ \citenamefont {Kieslich}}]{Vervoorts2019}%
  \BibitemOpen
  \bibfield  {author} {\bibinfo {author} {\bibfnamefont {P.}~\bibnamefont {Vervoorts}}, \bibinfo {author} {\bibfnamefont {C.~L.}\ \bibnamefont {Hobday}}, \bibinfo {author} {\bibfnamefont {M.~G.}\ \bibnamefont {Ehrenreich}}, \bibinfo {author} {\bibfnamefont {D.}~\bibnamefont {Daisenberger}}, \ and\ \bibinfo {author} {\bibfnamefont {G.}~\bibnamefont {Kieslich}},\ }\bibfield  {title} {\enquote {\bibinfo {title} {The zeolitic imidazolate framework zif‐4 under low hydrostatic pressures},}\ }\href {\doibase 10.1002/zaac.201900046} {\bibfield  {journal} {\bibinfo  {journal} {Zeitschrift f\"{u}r anorganische und allgemeine Chemie}\ }\textbf {\bibinfo {volume} {645}},\ \bibinfo {pages} {970–974} (\bibinfo {year} {2019})}\BibitemShut {NoStop}%
\bibitem [{\citenamefont {Widmer}\ \emph {et~al.}(2019{\natexlab{b}})\citenamefont {Widmer}, \citenamefont {Lampronti}, \citenamefont {Casati}, \citenamefont {Farsang}, \citenamefont {Bennett},\ and\ \citenamefont {Redfern}}]{Widmer2019_2}%
  \BibitemOpen
  \bibfield  {author} {\bibinfo {author} {\bibfnamefont {R.~N.}\ \bibnamefont {Widmer}}, \bibinfo {author} {\bibfnamefont {G.~I.}\ \bibnamefont {Lampronti}}, \bibinfo {author} {\bibfnamefont {N.}~\bibnamefont {Casati}}, \bibinfo {author} {\bibfnamefont {S.}~\bibnamefont {Farsang}}, \bibinfo {author} {\bibfnamefont {T.~D.}\ \bibnamefont {Bennett}}, \ and\ \bibinfo {author} {\bibfnamefont {S.~A.~T.}\ \bibnamefont {Redfern}},\ }\bibfield  {title} {\enquote {\bibinfo {title} {X-ray radiation-induced amorphization of metal{\textendash}organic frameworks},}\ }\href {\doibase 10.1039/c9cp01463b} {\bibfield  {journal} {\bibinfo  {journal} {Physical Chemistry Chemical Physics}\ }\textbf {\bibinfo {volume} {21}},\ \bibinfo {pages} {12389--12395} (\bibinfo {year} {2019}{\natexlab{b}})}\BibitemShut {NoStop}%
\bibitem [{\citenamefont {Beake}\ \emph {et~al.}(2013)\citenamefont {Beake}, \citenamefont {Dove}, \citenamefont {Phillips}, \citenamefont {Keen}, \citenamefont {Tucker}, \citenamefont {Goodwin}, \citenamefont {Bennett},\ and\ \citenamefont {Cheetham}}]{Beake2013}%
  \BibitemOpen
  \bibfield  {author} {\bibinfo {author} {\bibfnamefont {E.~O.~R.}\ \bibnamefont {Beake}}, \bibinfo {author} {\bibfnamefont {M.~T.}\ \bibnamefont {Dove}}, \bibinfo {author} {\bibfnamefont {A.~E.}\ \bibnamefont {Phillips}}, \bibinfo {author} {\bibfnamefont {D.~A.}\ \bibnamefont {Keen}}, \bibinfo {author} {\bibfnamefont {M.~G.}\ \bibnamefont {Tucker}}, \bibinfo {author} {\bibfnamefont {A.~L.}\ \bibnamefont {Goodwin}}, \bibinfo {author} {\bibfnamefont {T.~D.}\ \bibnamefont {Bennett}}, \ and\ \bibinfo {author} {\bibfnamefont {A.~K.}\ \bibnamefont {Cheetham}},\ }\bibfield  {title} {\enquote {\bibinfo {title} {Flexibility of zeolitic imidazolate framework structures studied by neutron total scattering and the reverse monte carlo method},}\ }\href {\doibase 10.1088/0953-8984/25/39/395403} {\bibfield  {journal} {\bibinfo  {journal} {Journal of Physics: Condensed Matter}\ }\textbf {\bibinfo {volume} {25}},\ \bibinfo {pages} {395403} (\bibinfo {year} {2013})}\BibitemShut {NoStop}%
\bibitem [{\citenamefont {Ryder}\ \emph {et~al.}(2014)\citenamefont {Ryder}, \citenamefont {Civalleri}, \citenamefont {Bennett}, \citenamefont {Henke}, \citenamefont {Rudić}, \citenamefont {Cinque}, \citenamefont {Fernandez-Alonso},\ and\ \citenamefont {Tan}}]{Ryder2014}%
  \BibitemOpen
  \bibfield  {author} {\bibinfo {author} {\bibfnamefont {M.}~\bibnamefont {Ryder}}, \bibinfo {author} {\bibfnamefont {B.}~\bibnamefont {Civalleri}}, \bibinfo {author} {\bibfnamefont {T.}~\bibnamefont {Bennett}}, \bibinfo {author} {\bibfnamefont {S.}~\bibnamefont {Henke}}, \bibinfo {author} {\bibfnamefont {S.}~\bibnamefont {Rudić}}, \bibinfo {author} {\bibfnamefont {G.}~\bibnamefont {Cinque}}, \bibinfo {author} {\bibfnamefont {F.}~\bibnamefont {Fernandez-Alonso}}, \ and\ \bibinfo {author} {\bibfnamefont {J.-C.}\ \bibnamefont {Tan}},\ }\bibfield  {title} {\enquote {\bibinfo {title} {Identifying the role of terahertz vibrations in metal-organic frameworks: From gate-opening phenomenon to shear-driven structural destabilization},}\ }\href {\doibase 10.1103/physrevlett.113.215502} {\bibfield  {journal} {\bibinfo  {journal} {Physical Review Letters}\ }\textbf {\bibinfo {volume} {113}} (\bibinfo {year} {2014}),\ 10.1103/physrevlett.113.215502}\BibitemShut {NoStop}%
\bibitem [{\citenamefont {Butler}\ \emph {et~al.}(2019)\citenamefont {Butler}, \citenamefont {Vervoorts}, \citenamefont {Ehrenreich}, \citenamefont {Armstrong}, \citenamefont {Skelton},\ and\ \citenamefont {Kieslich}}]{Butler2019}%
  \BibitemOpen
  \bibfield  {author} {\bibinfo {author} {\bibfnamefont {K.~T.}\ \bibnamefont {Butler}}, \bibinfo {author} {\bibfnamefont {P.}~\bibnamefont {Vervoorts}}, \bibinfo {author} {\bibfnamefont {M.~G.}\ \bibnamefont {Ehrenreich}}, \bibinfo {author} {\bibfnamefont {J.}~\bibnamefont {Armstrong}}, \bibinfo {author} {\bibfnamefont {J.~M.}\ \bibnamefont {Skelton}}, \ and\ \bibinfo {author} {\bibfnamefont {G.}~\bibnamefont {Kieslich}},\ }\bibfield  {title} {\enquote {\bibinfo {title} {Experimental evidence for vibrational entropy as driving parameter of flexibility in the metal–organic framework zif-4(zn)},}\ }\href {\doibase 10.1021/acs.chemmater.9b01908} {\bibfield  {journal} {\bibinfo  {journal} {Chemistry of Materials}\ }\textbf {\bibinfo {volume} {31}},\ \bibinfo {pages} {8366–8372} (\bibinfo {year} {2019})}\BibitemShut {NoStop}%
\bibitem [{\citenamefont {Tan}\ \emph {et~al.}(2015)\citenamefont {Tan}, \citenamefont {Civalleri}, \citenamefont {Erba},\ and\ \citenamefont {Albanese}}]{Tan2015}%
  \BibitemOpen
  \bibfield  {author} {\bibinfo {author} {\bibfnamefont {J.-C.}\ \bibnamefont {Tan}}, \bibinfo {author} {\bibfnamefont {B.}~\bibnamefont {Civalleri}}, \bibinfo {author} {\bibfnamefont {A.}~\bibnamefont {Erba}}, \ and\ \bibinfo {author} {\bibfnamefont {E.}~\bibnamefont {Albanese}},\ }\bibfield  {title} {\enquote {\bibinfo {title} {Quantum mechanical predictions to elucidate the anisotropic elastic properties of zeolitic imidazolate frameworks: Zif-4 vs. zif-zni},}\ }\href {\doibase 10.1039/c4ce01564a} {\bibfield  {journal} {\bibinfo  {journal} {CrystEngComm}\ }\textbf {\bibinfo {volume} {17}},\ \bibinfo {pages} {375–382} (\bibinfo {year} {2015})}\BibitemShut {NoStop}%
\bibitem [{\citenamefont {Ryder}\ and\ \citenamefont {Tan}(2016)}]{Ryder2016}%
  \BibitemOpen
  \bibfield  {author} {\bibinfo {author} {\bibfnamefont {M.~R.}\ \bibnamefont {Ryder}}\ and\ \bibinfo {author} {\bibfnamefont {J.-C.}\ \bibnamefont {Tan}},\ }\bibfield  {title} {\enquote {\bibinfo {title} {Explaining the mechanical mechanisms of zeolitic metal–organic frameworks: revealing auxeticity and anomalous elasticity},}\ }\href {\doibase 10.1039/c5dt03514g} {\bibfield  {journal} {\bibinfo  {journal} {Dalton Trans.}\ }\textbf {\bibinfo {volume} {45}},\ \bibinfo {pages} {4154–4161} (\bibinfo {year} {2016})}\BibitemShut {NoStop}%
\bibitem [{\citenamefont {Bouëssel~du Bourg}\ \emph {et~al.}(2014)\citenamefont {Bouëssel~du Bourg}, \citenamefont {Ortiz}, \citenamefont {Boutin},\ and\ \citenamefont {Coudert}}]{BousselduBourg2014}%
  \BibitemOpen
  \bibfield  {author} {\bibinfo {author} {\bibfnamefont {L.}~\bibnamefont {Bouëssel~du Bourg}}, \bibinfo {author} {\bibfnamefont {A.~U.}\ \bibnamefont {Ortiz}}, \bibinfo {author} {\bibfnamefont {A.}~\bibnamefont {Boutin}}, \ and\ \bibinfo {author} {\bibfnamefont {F.-X.}\ \bibnamefont {Coudert}},\ }\bibfield  {title} {\enquote {\bibinfo {title} {Thermal and mechanical stability of zeolitic imidazolate frameworks polymorphs},}\ }\href {\doibase 10.1063/1.4904818} {\bibfield  {journal} {\bibinfo  {journal} {APL Materials}\ }\textbf {\bibinfo {volume} {2}} (\bibinfo {year} {2014}),\ 10.1063/1.4904818}\BibitemShut {NoStop}%
\bibitem [{\citenamefont {Castel}\ and\ \citenamefont {Coudert}(2023)}]{Castel2023_2}%
  \BibitemOpen
  \bibfield  {author} {\bibinfo {author} {\bibfnamefont {N.}~\bibnamefont {Castel}}\ and\ \bibinfo {author} {\bibfnamefont {F.-X.}\ \bibnamefont {Coudert}},\ }\bibfield  {title} {\enquote {\bibinfo {title} {Computation of finite temperature mechanical properties of zeolitic imidazolate framework glasses by molecular dynamics},}\ }\href {\doibase 10.1021/acs.chemmater.3c00392} {\bibfield  {journal} {\bibinfo  {journal} {Chemistry of Materials}\ }\textbf {\bibinfo {volume} {35}},\ \bibinfo {pages} {4038–4047} (\bibinfo {year} {2023})}\BibitemShut {NoStop}%
\bibitem [{\citenamefont {Zhang}\ \emph {et~al.}(2019)\citenamefont {Zhang}, \citenamefont {Qiao}, \citenamefont {Tao},\ and\ \citenamefont {Yue}}]{Zhang2019}%
  \BibitemOpen
  \bibfield  {author} {\bibinfo {author} {\bibfnamefont {J.}~\bibnamefont {Zhang}}, \bibinfo {author} {\bibfnamefont {A.}~\bibnamefont {Qiao}}, \bibinfo {author} {\bibfnamefont {H.}~\bibnamefont {Tao}}, \ and\ \bibinfo {author} {\bibfnamefont {Y.}~\bibnamefont {Yue}},\ }\bibfield  {title} {\enquote {\bibinfo {title} {Synthesis, phase transitions and vitrification of the zeolitic imidazolate framework: Zif-4},}\ }\href {\doibase 10.1016/j.jnoncrysol.2019.119665} {\bibfield  {journal} {\bibinfo  {journal} {Journal of Non-Crystalline Solids}\ }\textbf {\bibinfo {volume} {525}},\ \bibinfo {pages} {119665} (\bibinfo {year} {2019})}\BibitemShut {NoStop}%
\bibitem [{\citenamefont {Gong}\ \emph {et~al.}(2022)\citenamefont {Gong}, \citenamefont {Gnanasekaran}, \citenamefont {Ma}, \citenamefont {Forman}, \citenamefont {Wang}, \citenamefont {Su}, \citenamefont {Farha},\ and\ \citenamefont {Gianneschi}}]{Gong2022}%
  \BibitemOpen
  \bibfield  {author} {\bibinfo {author} {\bibfnamefont {X.}~\bibnamefont {Gong}}, \bibinfo {author} {\bibfnamefont {K.}~\bibnamefont {Gnanasekaran}}, \bibinfo {author} {\bibfnamefont {K.}~\bibnamefont {Ma}}, \bibinfo {author} {\bibfnamefont {C.~J.}\ \bibnamefont {Forman}}, \bibinfo {author} {\bibfnamefont {X.}~\bibnamefont {Wang}}, \bibinfo {author} {\bibfnamefont {S.}~\bibnamefont {Su}}, \bibinfo {author} {\bibfnamefont {O.~K.}\ \bibnamefont {Farha}}, \ and\ \bibinfo {author} {\bibfnamefont {N.~C.}\ \bibnamefont {Gianneschi}},\ }\bibfield  {title} {\enquote {\bibinfo {title} {Rapid generation of metal–organic framework phase diagrams by high-throughput transmission electron microscopy},}\ }\href {\doibase 10.1021/jacs.2c01095} {\bibfield  {journal} {\bibinfo  {journal} {Journal of the American Chemical Society}\ }\textbf {\bibinfo {volume} {144}},\ \bibinfo {pages} {6674–6680} (\bibinfo {year} {2022})}\BibitemShut {NoStop}%
\bibitem [{\citenamefont {Widmer}\ \emph {et~al.}(2019{\natexlab{c}})\citenamefont {Widmer}, \citenamefont {Lampronti}, \citenamefont {Anzellini}, \citenamefont {Gaillac}, \citenamefont {Farsang}, \citenamefont {Zhou}, \citenamefont {Belenguer}, \citenamefont {Wilson}, \citenamefont {Palmer}, \citenamefont {Kleppe}, \citenamefont {Wharmby}, \citenamefont {Yu}, \citenamefont {Cohen}, \citenamefont {Telfer}, \citenamefont {Redfern}, \citenamefont {Coudert}, \citenamefont {MacLeod},\ and\ \citenamefont {Bennett}}]{Widmer2019Nature}%
  \BibitemOpen
  \bibfield  {author} {\bibinfo {author} {\bibfnamefont {R.~N.}\ \bibnamefont {Widmer}}, \bibinfo {author} {\bibfnamefont {G.~I.}\ \bibnamefont {Lampronti}}, \bibinfo {author} {\bibfnamefont {S.}~\bibnamefont {Anzellini}}, \bibinfo {author} {\bibfnamefont {R.}~\bibnamefont {Gaillac}}, \bibinfo {author} {\bibfnamefont {S.}~\bibnamefont {Farsang}}, \bibinfo {author} {\bibfnamefont {C.}~\bibnamefont {Zhou}}, \bibinfo {author} {\bibfnamefont {A.~M.}\ \bibnamefont {Belenguer}}, \bibinfo {author} {\bibfnamefont {C.~W.}\ \bibnamefont {Wilson}}, \bibinfo {author} {\bibfnamefont {H.}~\bibnamefont {Palmer}}, \bibinfo {author} {\bibfnamefont {A.~K.}\ \bibnamefont {Kleppe}}, \bibinfo {author} {\bibfnamefont {M.~T.}\ \bibnamefont {Wharmby}}, \bibinfo {author} {\bibfnamefont {X.}~\bibnamefont {Yu}}, \bibinfo {author} {\bibfnamefont {S.~M.}\ \bibnamefont {Cohen}}, \bibinfo {author} {\bibfnamefont {S.~G.}\ \bibnamefont {Telfer}}, \bibinfo {author} {\bibfnamefont {S.~A.~T.}\ \bibnamefont {Redfern}}, \bibinfo {author}
  {\bibfnamefont {F.-X.}\ \bibnamefont {Coudert}}, \bibinfo {author} {\bibfnamefont {S.~G.}\ \bibnamefont {MacLeod}}, \ and\ \bibinfo {author} {\bibfnamefont {T.~D.}\ \bibnamefont {Bennett}},\ }\bibfield  {title} {\enquote {\bibinfo {title} {Pressure promoted low-temperature melting of metal–organic frameworks},}\ }\href {\doibase 10.1038/s41563-019-0317-4} {\bibfield  {journal} {\bibinfo  {journal} {Nature Materials}\ }\textbf {\bibinfo {volume} {18}},\ \bibinfo {pages} {370–376} (\bibinfo {year} {2019}{\natexlab{c}})}\BibitemShut {NoStop}%
\bibitem [{\citenamefont {Méndez}\ and\ \citenamefont {Semino}(2024)}]{Mendez2024}%
  \BibitemOpen
  \bibfield  {author} {\bibinfo {author} {\bibfnamefont {E.}~\bibnamefont {Méndez}}\ and\ \bibinfo {author} {\bibfnamefont {R.}~\bibnamefont {Semino}},\ }\bibfield  {title} {\enquote {\bibinfo {title} {Microscopic mechanism of thermal amorphization of zif-4 and melting of zif-zni revealed via molecular dynamics and machine learning techniques},}\ }\href {\doibase 10.1039/d3ta07361k} {\bibfield  {journal} {\bibinfo  {journal} {Journal of Materials Chemistry A}\ }\textbf {\bibinfo {volume} {12}},\ \bibinfo {pages} {4572–4582} (\bibinfo {year} {2024})}\BibitemShut {NoStop}%
\bibitem [{\citenamefont {Balestra}\ and\ \citenamefont {Semino}(2022)}]{Balestra2022}%
  \BibitemOpen
  \bibfield  {author} {\bibinfo {author} {\bibfnamefont {S.~R.~G.}\ \bibnamefont {Balestra}}\ and\ \bibinfo {author} {\bibfnamefont {R.}~\bibnamefont {Semino}},\ }\bibfield  {title} {\enquote {\bibinfo {title} {Computer simulation of the early stages of self-assembly and thermal decomposition of {ZIF}-8},}\ }\href {\doibase 10.1063/5.0128656} {\bibfield  {journal} {\bibinfo  {journal} {The Journal of Chemical Physics}\ }\textbf {\bibinfo {volume} {157}},\ \bibinfo {pages} {184502} (\bibinfo {year} {2022})}\BibitemShut {NoStop}%
\bibitem [{\citenamefont {Shi}\ \emph {et~al.}(2024)\citenamefont {Shi}, \citenamefont {Liu}, \citenamefont {Yue}, \citenamefont {Arramel},\ and\ \citenamefont {Li}}]{Shi2024}%
  \BibitemOpen
  \bibfield  {author} {\bibinfo {author} {\bibfnamefont {Z.}~\bibnamefont {Shi}}, \bibinfo {author} {\bibfnamefont {B.}~\bibnamefont {Liu}}, \bibinfo {author} {\bibfnamefont {Y.}~\bibnamefont {Yue}}, \bibinfo {author} {\bibfnamefont {A.}~\bibnamefont {Arramel}}, \ and\ \bibinfo {author} {\bibfnamefont {N.}~\bibnamefont {Li}},\ }\bibfield  {title} {\enquote {\bibinfo {title} {Unraveling medium‐range order and melting mechanism of zif‐4 under high temperature},}\ }\href {\doibase 10.1111/jace.19741} {\bibfield  {journal} {\bibinfo  {journal} {Journal of the American Ceramic Society}\ }\textbf {\bibinfo {volume} {107}},\ \bibinfo {pages} {3845–3856} (\bibinfo {year} {2024})}\BibitemShut {NoStop}%
\bibitem [{\citenamefont {Du}\ \emph {et~al.}(2024)\citenamefont {Du}, \citenamefont {Li}, \citenamefont {Ganisetti}, \citenamefont {Bauchy}, \citenamefont {Yue},\ and\ \citenamefont {Smedskjaer}}]{Du2024}%
  \BibitemOpen
  \bibfield  {author} {\bibinfo {author} {\bibfnamefont {T.}~\bibnamefont {Du}}, \bibinfo {author} {\bibfnamefont {S.}~\bibnamefont {Li}}, \bibinfo {author} {\bibfnamefont {S.}~\bibnamefont {Ganisetti}}, \bibinfo {author} {\bibfnamefont {M.}~\bibnamefont {Bauchy}}, \bibinfo {author} {\bibfnamefont {Y.}~\bibnamefont {Yue}}, \ and\ \bibinfo {author} {\bibfnamefont {M.~M.}\ \bibnamefont {Smedskjaer}},\ }\bibfield  {title} {\enquote {\bibinfo {title} {Deciphering the controlling factors for phase transitions in zeolitic imidazolate frameworks},}\ }\href {\doibase 10.1093/nsr/nwae023} {\bibfield  {journal} {\bibinfo  {journal} {National Science Review}\ }\textbf {\bibinfo {volume} {11}} (\bibinfo {year} {2024}),\ 10.1093/nsr/nwae023}\BibitemShut {NoStop}%
\bibitem [{\citenamefont {Castel}\ \emph {et~al.}(2024)\citenamefont {Castel}, \citenamefont {André}, \citenamefont {Edwards}, \citenamefont {Evans},\ and\ \citenamefont {Coudert}}]{Castel2024}%
  \BibitemOpen
  \bibfield  {author} {\bibinfo {author} {\bibfnamefont {N.}~\bibnamefont {Castel}}, \bibinfo {author} {\bibfnamefont {D.}~\bibnamefont {André}}, \bibinfo {author} {\bibfnamefont {C.}~\bibnamefont {Edwards}}, \bibinfo {author} {\bibfnamefont {J.~D.}\ \bibnamefont {Evans}}, \ and\ \bibinfo {author} {\bibfnamefont {F.-X.}\ \bibnamefont {Coudert}},\ }\bibfield  {title} {\enquote {\bibinfo {title} {Machine learning interatomic potentials for amorphous zeolitic imidazolate frameworks},}\ }\href {\doibase 10.1039/d3dd00236e} {\bibfield  {journal} {\bibinfo  {journal} {Digital Discovery}\ }\textbf {\bibinfo {volume} {3}},\ \bibinfo {pages} {355–368} (\bibinfo {year} {2024})}\BibitemShut {NoStop}%
\bibitem [{\citenamefont {Gaillac}, \citenamefont {Pullumbi},\ and\ \citenamefont {Coudert}(2018)}]{Gaillac2018}%
  \BibitemOpen
  \bibfield  {author} {\bibinfo {author} {\bibfnamefont {R.}~\bibnamefont {Gaillac}}, \bibinfo {author} {\bibfnamefont {P.}~\bibnamefont {Pullumbi}}, \ and\ \bibinfo {author} {\bibfnamefont {F.-X.}\ \bibnamefont {Coudert}},\ }\bibfield  {title} {\enquote {\bibinfo {title} {Melting of zeolitic imidazolate frameworks with different topologies: Insight from first-principles molecular dynamics},}\ }\href {\doibase 10.1021/acs.jpcc.8b00385} {\bibfield  {journal} {\bibinfo  {journal} {The Journal of Physical Chemistry C}\ }\textbf {\bibinfo {volume} {122}},\ \bibinfo {pages} {6730--6736} (\bibinfo {year} {2018})}\BibitemShut {NoStop}%
\bibitem [{\citenamefont {Yang}\ \emph {et~al.}(2018)\citenamefont {Yang}, \citenamefont {Shin}, \citenamefont {Li}, \citenamefont {Bennett}, \citenamefont {van Duin},\ and\ \citenamefont {Mauro}}]{Yang2018}%
  \BibitemOpen
  \bibfield  {author} {\bibinfo {author} {\bibfnamefont {Y.}~\bibnamefont {Yang}}, \bibinfo {author} {\bibfnamefont {Y.~K.}\ \bibnamefont {Shin}}, \bibinfo {author} {\bibfnamefont {S.}~\bibnamefont {Li}}, \bibinfo {author} {\bibfnamefont {T.~D.}\ \bibnamefont {Bennett}}, \bibinfo {author} {\bibfnamefont {A.~C.~T.}\ \bibnamefont {van Duin}}, \ and\ \bibinfo {author} {\bibfnamefont {J.~C.}\ \bibnamefont {Mauro}},\ }\bibfield  {title} {\enquote {\bibinfo {title} {Enabling computational design of zifs using reaxff},}\ }\href {\doibase 10.1021/acs.jpcb.8b08094} {\bibfield  {journal} {\bibinfo  {journal} {The Journal of Physical Chemistry B}\ }\textbf {\bibinfo {volume} {122}},\ \bibinfo {pages} {9616–9624} (\bibinfo {year} {2018})}\BibitemShut {NoStop}%
\bibitem [{\citenamefont {Castel}\ and\ \citenamefont {Coudert}(2022)}]{Castel2022}%
  \BibitemOpen
  \bibfield  {author} {\bibinfo {author} {\bibfnamefont {N.}~\bibnamefont {Castel}}\ and\ \bibinfo {author} {\bibfnamefont {F.-X.}\ \bibnamefont {Coudert}},\ }\bibfield  {title} {\enquote {\bibinfo {title} {Challenges in molecular dynamics of amorphous {ZIFs} using reactive force fields},}\ }\href {\doibase 10.1021/acs.jpcc.2c06305} {\bibfield  {journal} {\bibinfo  {journal} {The Journal of Physical Chemistry C}\ }\textbf {\bibinfo {volume} {126}},\ \bibinfo {pages} {19532--19541} (\bibinfo {year} {2022})}\BibitemShut {NoStop}%
\bibitem [{\citenamefont {Sapnik}\ \emph {et~al.}(2021)\citenamefont {Sapnik}, \citenamefont {Bechis}, \citenamefont {Collins}, \citenamefont {Johnstone}, \citenamefont {Divitini}, \citenamefont {Smith}, \citenamefont {Chater}, \citenamefont {Addicoat}, \citenamefont {Johnson}, \citenamefont {Keen}, \citenamefont {Jelfs},\ and\ \citenamefont {Bennett}}]{Sapnik2021}%
  \BibitemOpen
  \bibfield  {author} {\bibinfo {author} {\bibfnamefont {A.~F.}\ \bibnamefont {Sapnik}}, \bibinfo {author} {\bibfnamefont {I.}~\bibnamefont {Bechis}}, \bibinfo {author} {\bibfnamefont {S.~M.}\ \bibnamefont {Collins}}, \bibinfo {author} {\bibfnamefont {D.~N.}\ \bibnamefont {Johnstone}}, \bibinfo {author} {\bibfnamefont {G.}~\bibnamefont {Divitini}}, \bibinfo {author} {\bibfnamefont {A.~J.}\ \bibnamefont {Smith}}, \bibinfo {author} {\bibfnamefont {P.~A.}\ \bibnamefont {Chater}}, \bibinfo {author} {\bibfnamefont {M.~A.}\ \bibnamefont {Addicoat}}, \bibinfo {author} {\bibfnamefont {T.}~\bibnamefont {Johnson}}, \bibinfo {author} {\bibfnamefont {D.~A.}\ \bibnamefont {Keen}}, \bibinfo {author} {\bibfnamefont {K.~E.}\ \bibnamefont {Jelfs}}, \ and\ \bibinfo {author} {\bibfnamefont {T.~D.}\ \bibnamefont {Bennett}},\ }\bibfield  {title} {\enquote {\bibinfo {title} {Mixed hierarchical local structure in a disordered metal–organic framework},}\ }\href {\doibase 10.1038/s41467-021-22218-9} {\bibfield  {journal} {\bibinfo
  {journal} {Nature Communications}\ }\textbf {\bibinfo {volume} {12}} (\bibinfo {year} {2021}),\ 10.1038/s41467-021-22218-9}\BibitemShut {NoStop}%
\bibitem [{\citenamefont {Stracke}\ and\ \citenamefont {Evans}(2024)}]{Stracke2024}%
  \BibitemOpen
  \bibfield  {author} {\bibinfo {author} {\bibfnamefont {K.}~\bibnamefont {Stracke}}\ and\ \bibinfo {author} {\bibfnamefont {J.~D.}\ \bibnamefont {Evans}},\ }\bibfield  {title} {\enquote {\bibinfo {title} {The use of collective variables and enhanced sampling in the simulations of existing and emerging microporous materials},}\ }\href {\doibase 10.1039/d4nr01024h} {\bibfield  {journal} {\bibinfo  {journal} {Nanoscale}\ } (\bibinfo {year} {2024}),\ 10.1039/d4nr01024h}\BibitemShut {NoStop}%
\bibitem [{\citenamefont {Laio}\ and\ \citenamefont {Parrinello}(2002)}]{Laio2002}%
  \BibitemOpen
  \bibfield  {author} {\bibinfo {author} {\bibfnamefont {A.}~\bibnamefont {Laio}}\ and\ \bibinfo {author} {\bibfnamefont {M.}~\bibnamefont {Parrinello}},\ }\bibfield  {title} {\enquote {\bibinfo {title} {Escaping free-energy minima},}\ }\href {\doibase 10.1073/pnas.202427399} {\bibfield  {journal} {\bibinfo  {journal} {Proceedings of the National Academy of Sciences}\ }\textbf {\bibinfo {volume} {99}},\ \bibinfo {pages} {12562–12566} (\bibinfo {year} {2002})}\BibitemShut {NoStop}%
\bibitem [{\citenamefont {Sch\"{a}fer}\ and\ \citenamefont {Settanni}(2020)}]{Schfer2020}%
  \BibitemOpen
  \bibfield  {author} {\bibinfo {author} {\bibfnamefont {T.~M.}\ \bibnamefont {Sch\"{a}fer}}\ and\ \bibinfo {author} {\bibfnamefont {G.}~\bibnamefont {Settanni}},\ }\bibfield  {title} {\enquote {\bibinfo {title} {Data reweighting in metadynamics simulations},}\ }\href {\doibase 10.1021/acs.jctc.9b00867} {\bibfield  {journal} {\bibinfo  {journal} {Journal of Chemical Theory and Computation}\ }\textbf {\bibinfo {volume} {16}},\ \bibinfo {pages} {2042–2052} (\bibinfo {year} {2020})}\BibitemShut {NoStop}%
\bibitem [{\citenamefont {Barducci}, \citenamefont {Bussi},\ and\ \citenamefont {Parrinello}(2008)}]{Barducci2008}%
  \BibitemOpen
  \bibfield  {author} {\bibinfo {author} {\bibfnamefont {A.}~\bibnamefont {Barducci}}, \bibinfo {author} {\bibfnamefont {G.}~\bibnamefont {Bussi}}, \ and\ \bibinfo {author} {\bibfnamefont {M.}~\bibnamefont {Parrinello}},\ }\bibfield  {title} {\enquote {\bibinfo {title} {Well-tempered metadynamics: A smoothly converging and tunable free-energy method},}\ }\href {\doibase 10.1103/physrevlett.100.020603} {\bibfield  {journal} {\bibinfo  {journal} {Physical Review Letters}\ }\textbf {\bibinfo {volume} {100}} (\bibinfo {year} {2008}),\ 10.1103/physrevlett.100.020603}\BibitemShut {NoStop}%
\bibitem [{\citenamefont {Bussi}\ and\ \citenamefont {Laio}(2020)}]{Bussi2020}%
  \BibitemOpen
  \bibfield  {author} {\bibinfo {author} {\bibfnamefont {G.}~\bibnamefont {Bussi}}\ and\ \bibinfo {author} {\bibfnamefont {A.}~\bibnamefont {Laio}},\ }\bibfield  {title} {\enquote {\bibinfo {title} {Using metadynamics to explore complex free-energy landscapes},}\ }\href {\doibase 10.1038/s42254-020-0153-0} {\bibfield  {journal} {\bibinfo  {journal} {Nature Reviews Physics}\ }\textbf {\bibinfo {volume} {2}},\ \bibinfo {pages} {200–212} (\bibinfo {year} {2020})}\BibitemShut {NoStop}%
\bibitem [{\citenamefont {Piaggi}\ and\ \citenamefont {Parrinello}(2019)}]{Piaggi2019}%
  \BibitemOpen
  \bibfield  {author} {\bibinfo {author} {\bibfnamefont {P.~M.}\ \bibnamefont {Piaggi}}\ and\ \bibinfo {author} {\bibfnamefont {M.}~\bibnamefont {Parrinello}},\ }\bibfield  {title} {\enquote {\bibinfo {title} {Calculation of phase diagrams in the multithermal-multibaric ensemble},}\ }\href {\doibase 10.1063/1.5102104} {\bibfield  {journal} {\bibinfo  {journal} {The Journal of Chemical Physics}\ }\textbf {\bibinfo {volume} {150}} (\bibinfo {year} {2019}),\ 10.1063/1.5102104}\BibitemShut {NoStop}%
\bibitem [{\citenamefont {Martoňák}, \citenamefont {Laio},\ and\ \citenamefont {Parrinello}(2003)}]{Martok2003}%
  \BibitemOpen
  \bibfield  {author} {\bibinfo {author} {\bibfnamefont {R.}~\bibnamefont {Martoňák}}, \bibinfo {author} {\bibfnamefont {A.}~\bibnamefont {Laio}}, \ and\ \bibinfo {author} {\bibfnamefont {M.}~\bibnamefont {Parrinello}},\ }\bibfield  {title} {\enquote {\bibinfo {title} {Predicting crystal structures: The parrinello-rahman method revisited},}\ }\href {\doibase 10.1103/physrevlett.90.075503} {\bibfield  {journal} {\bibinfo  {journal} {Physical Review Letters}\ }\textbf {\bibinfo {volume} {90}} (\bibinfo {year} {2003}),\ 10.1103/physrevlett.90.075503}\BibitemShut {NoStop}%
\bibitem [{\citenamefont {Thompson}\ \emph {et~al.}(2022)\citenamefont {Thompson}, \citenamefont {Aktulga}, \citenamefont {Berger}, \citenamefont {Bolintineanu}, \citenamefont {Brown}, \citenamefont {Crozier}, \citenamefont {in~{\textquotesingle}t~Veld}, \citenamefont {Kohlmeyer}, \citenamefont {Moore}, \citenamefont {Nguyen}, \citenamefont {Shan}, \citenamefont {Stevens}, \citenamefont {Tranchida}, \citenamefont {Trott},\ and\ \citenamefont {Plimpton}}]{lammps}%
  \BibitemOpen
  \bibfield  {author} {\bibinfo {author} {\bibfnamefont {A.~P.}\ \bibnamefont {Thompson}}, \bibinfo {author} {\bibfnamefont {H.~M.}\ \bibnamefont {Aktulga}}, \bibinfo {author} {\bibfnamefont {R.}~\bibnamefont {Berger}}, \bibinfo {author} {\bibfnamefont {D.~S.}\ \bibnamefont {Bolintineanu}}, \bibinfo {author} {\bibfnamefont {W.~M.}\ \bibnamefont {Brown}}, \bibinfo {author} {\bibfnamefont {P.~S.}\ \bibnamefont {Crozier}}, \bibinfo {author} {\bibfnamefont {P.~J.}\ \bibnamefont {in~{\textquotesingle}t~Veld}}, \bibinfo {author} {\bibfnamefont {A.}~\bibnamefont {Kohlmeyer}}, \bibinfo {author} {\bibfnamefont {S.~G.}\ \bibnamefont {Moore}}, \bibinfo {author} {\bibfnamefont {T.~D.}\ \bibnamefont {Nguyen}}, \bibinfo {author} {\bibfnamefont {R.}~\bibnamefont {Shan}}, \bibinfo {author} {\bibfnamefont {M.~J.}\ \bibnamefont {Stevens}}, \bibinfo {author} {\bibfnamefont {J.}~\bibnamefont {Tranchida}}, \bibinfo {author} {\bibfnamefont {C.}~\bibnamefont {Trott}}, \ and\ \bibinfo {author} {\bibfnamefont {S.~J.}\ \bibnamefont
  {Plimpton}},\ }\bibfield  {title} {\enquote {\bibinfo {title} {{LAMMPS} - a flexible simulation tool for particle-based materials modeling at the atomic, meso, and continuum scales},}\ }\href {\doibase 10.1016/j.cpc.2021.108171} {\bibfield  {journal} {\bibinfo  {journal} {Computer Physics Communications}\ }\textbf {\bibinfo {volume} {271}},\ \bibinfo {pages} {108171} (\bibinfo {year} {2022})}\BibitemShut {NoStop}%
\bibitem [{\citenamefont {Tribello}\ \emph {et~al.}(2014)\citenamefont {Tribello}, \citenamefont {Bonomi}, \citenamefont {Branduardi}, \citenamefont {Camilloni},\ and\ \citenamefont {Bussi}}]{Tribello2014}%
  \BibitemOpen
  \bibfield  {author} {\bibinfo {author} {\bibfnamefont {G.~A.}\ \bibnamefont {Tribello}}, \bibinfo {author} {\bibfnamefont {M.}~\bibnamefont {Bonomi}}, \bibinfo {author} {\bibfnamefont {D.}~\bibnamefont {Branduardi}}, \bibinfo {author} {\bibfnamefont {C.}~\bibnamefont {Camilloni}}, \ and\ \bibinfo {author} {\bibfnamefont {G.}~\bibnamefont {Bussi}},\ }\bibfield  {title} {\enquote {\bibinfo {title} {Plumed 2: New feathers for an old bird},}\ }\href {\doibase 10.1016/j.cpc.2013.09.018} {\bibfield  {journal} {\bibinfo  {journal} {Computer Physics Communications}\ }\textbf {\bibinfo {volume} {185}},\ \bibinfo {pages} {604–613} (\bibinfo {year} {2014})}\BibitemShut {NoStop}%
\bibitem [{\citenamefont {Weng}\ and\ \citenamefont {Schmidt}(2019)}]{Weng2019}%
  \BibitemOpen
  \bibfield  {author} {\bibinfo {author} {\bibfnamefont {T.}~\bibnamefont {Weng}}\ and\ \bibinfo {author} {\bibfnamefont {J.~R.}\ \bibnamefont {Schmidt}},\ }\bibfield  {title} {\enquote {\bibinfo {title} {Flexible and transferable ab initio force field for zeolitic imidazolate frameworks: Zif-ff},}\ }\href {\doibase 10.1021/acs.jpca.8b12311} {\bibfield  {journal} {\bibinfo  {journal} {The Journal of Physical Chemistry A}\ }\textbf {\bibinfo {volume} {123}},\ \bibinfo {pages} {3000–3012} (\bibinfo {year} {2019})}\BibitemShut {NoStop}%
\bibitem [{\citenamefont {Evans}\ and\ \citenamefont {Holian}(1985)}]{Evans1985}%
  \BibitemOpen
  \bibfield  {author} {\bibinfo {author} {\bibfnamefont {D.~J.}\ \bibnamefont {Evans}}\ and\ \bibinfo {author} {\bibfnamefont {B.~L.}\ \bibnamefont {Holian}},\ }\bibfield  {title} {\enquote {\bibinfo {title} {The nose–hoover thermostat},}\ }\href {\doibase 10.1063/1.449071} {\bibfield  {journal} {\bibinfo  {journal} {The Journal of Chemical Physics}\ }\textbf {\bibinfo {volume} {83}},\ \bibinfo {pages} {4069–4074} (\bibinfo {year} {1985})}\BibitemShut {NoStop}%
\bibitem [{\citenamefont {Raiteri}\ \emph {et~al.}(2005)\citenamefont {Raiteri}, \citenamefont {Laio}, \citenamefont {Gervasio}, \citenamefont {Micheletti},\ and\ \citenamefont {Parrinello}}]{Raiteri2005}%
  \BibitemOpen
  \bibfield  {author} {\bibinfo {author} {\bibfnamefont {P.}~\bibnamefont {Raiteri}}, \bibinfo {author} {\bibfnamefont {A.}~\bibnamefont {Laio}}, \bibinfo {author} {\bibfnamefont {F.~L.}\ \bibnamefont {Gervasio}}, \bibinfo {author} {\bibfnamefont {C.}~\bibnamefont {Micheletti}}, \ and\ \bibinfo {author} {\bibfnamefont {M.}~\bibnamefont {Parrinello}},\ }\bibfield  {title} {\enquote {\bibinfo {title} {Efficient reconstruction of complex free energy landscapes by multiple walkers metadynamics},}\ }\href {\doibase 10.1021/jp054359r} {\bibfield  {journal} {\bibinfo  {journal} {The Journal of Physical Chemistry B}\ }\textbf {\bibinfo {volume} {110}},\ \bibinfo {pages} {3533–3539} (\bibinfo {year} {2005})}\BibitemShut {NoStop}%
\bibitem [{\citenamefont {Tiwary}\ and\ \citenamefont {Parrinello}(2014)}]{Tiwary2014}%
  \BibitemOpen
  \bibfield  {author} {\bibinfo {author} {\bibfnamefont {P.}~\bibnamefont {Tiwary}}\ and\ \bibinfo {author} {\bibfnamefont {M.}~\bibnamefont {Parrinello}},\ }\bibfield  {title} {\enquote {\bibinfo {title} {A time-independent free energy estimator for metadynamics},}\ }\href {\doibase 10.1021/jp504920s} {\bibfield  {journal} {\bibinfo  {journal} {The Journal of Physical Chemistry B}\ }\textbf {\bibinfo {volume} {119}},\ \bibinfo {pages} {736–742} (\bibinfo {year} {2014})}\BibitemShut {NoStop}%
\bibitem [{\citenamefont {Wharmby}\ \emph {et~al.}(2015)\citenamefont {Wharmby}, \citenamefont {Henke}, \citenamefont {Bennett}, \citenamefont {Bajpe}, \citenamefont {Schwedler}, \citenamefont {Thompson}, \citenamefont {Gozzo}, \citenamefont {Simoncic}, \citenamefont {Mellot‐Draznieks}, \citenamefont {Tao}, \citenamefont {Yue},\ and\ \citenamefont {Cheetham}}]{Wharmby2015}%
  \BibitemOpen
  \bibfield  {author} {\bibinfo {author} {\bibfnamefont {M.~T.}\ \bibnamefont {Wharmby}}, \bibinfo {author} {\bibfnamefont {S.}~\bibnamefont {Henke}}, \bibinfo {author} {\bibfnamefont {T.~D.}\ \bibnamefont {Bennett}}, \bibinfo {author} {\bibfnamefont {S.~R.}\ \bibnamefont {Bajpe}}, \bibinfo {author} {\bibfnamefont {I.}~\bibnamefont {Schwedler}}, \bibinfo {author} {\bibfnamefont {S.~P.}\ \bibnamefont {Thompson}}, \bibinfo {author} {\bibfnamefont {F.}~\bibnamefont {Gozzo}}, \bibinfo {author} {\bibfnamefont {P.}~\bibnamefont {Simoncic}}, \bibinfo {author} {\bibfnamefont {C.}~\bibnamefont {Mellot‐Draznieks}}, \bibinfo {author} {\bibfnamefont {H.}~\bibnamefont {Tao}}, \bibinfo {author} {\bibfnamefont {Y.}~\bibnamefont {Yue}}, \ and\ \bibinfo {author} {\bibfnamefont {A.~K.}\ \bibnamefont {Cheetham}},\ }\bibfield  {title} {\enquote {\bibinfo {title} {Extreme flexibility in a zeolitic imidazolate framework: Porous to dense phase transition in desolvated zif‐4},}\ }\href {\doibase 10.1002/anie.201410167}
  {\bibfield  {journal} {\bibinfo  {journal} {Angewandte Chemie International Edition}\ }\textbf {\bibinfo {volume} {54}},\ \bibinfo {pages} {6447–6451} (\bibinfo {year} {2015})}\BibitemShut {NoStop}%
\bibitem [{\citenamefont {Rogge}, \citenamefont {Waroquier},\ and\ \citenamefont {Van~Speybroeck}(2019)}]{Rogge2019}%
  \BibitemOpen
  \bibfield  {author} {\bibinfo {author} {\bibfnamefont {S.~M.~J.}\ \bibnamefont {Rogge}}, \bibinfo {author} {\bibfnamefont {M.}~\bibnamefont {Waroquier}}, \ and\ \bibinfo {author} {\bibfnamefont {V.}~\bibnamefont {Van~Speybroeck}},\ }\bibfield  {title} {\enquote {\bibinfo {title} {Unraveling the thermodynamic criteria for size-dependent spontaneous phase separation in soft porous crystals},}\ }\href {\doibase 10.1038/s41467-019-12754-w} {\bibfield  {journal} {\bibinfo  {journal} {Nature Communications}\ }\textbf {\bibinfo {volume} {10}} (\bibinfo {year} {2019}),\ 10.1038/s41467-019-12754-w}\BibitemShut {NoStop}%
\bibitem [{\citenamefont {Bussi}\ and\ \citenamefont {Tribello}(2019)}]{Bussi2019}%
  \BibitemOpen
  \bibfield  {author} {\bibinfo {author} {\bibfnamefont {G.}~\bibnamefont {Bussi}}\ and\ \bibinfo {author} {\bibfnamefont {G.~A.}\ \bibnamefont {Tribello}},\ }\href {\doibase 10.1007/978-1-4939-9608-7_21} {\emph {\bibinfo {title} {Biomolecular Simulations}}}\ (\bibinfo  {publisher} {Springer New York},\ \bibinfo {year} {2019})\ p.\ \bibinfo {pages} {529–578}\BibitemShut {NoStop}%
\end{thebibliography}%

\end{document}